\documentclass[aps, pre, 12pt, superscriptaddress, amsmath, amssymb, tightenlines]{revtex4-2}

\usepackage{graphicx}
\usepackage{amsmath}
\usepackage{amssymb}
\usepackage{tensor}
\usepackage{xcolor}
\usepackage{hyperref}

\begin{document}

\title{
Negative radiation pressure in Bose--Einstein condensates
}
\author{Dominik Ciurla}
\email{dominik.ciurla@doctoral.uj.edu.pl}
\affiliation{Institute of Theoretical Physics, Jagiellonian University, {\L}ojasiewicza 11, 30-348 Krak\'ow, Poland}
\author{P\'eter Forg\'acs}
\affiliation{Wigner RCP RMI, H1525 Budapest, POB 49, Hungary}
\affiliation{Institut Denis-Poisson CNRS/UMR 7013, Université de Tours, Parc de
Grandmont, 37200 Tours, France}
\author{\'Arp\'ad Luk\'acs}
\affiliation{Durham University, Department of Mathematical Sciences, Stockton Road, Durham, DH1 3LE, United Kingdom}
\affiliation{Wigner RCP RMI, H1525 Budapest, POB 49, Hungary}
\author{Tomasz Roma\'nczukiewicz}
\affiliation{Institute of Theoretical Physics, Jagiellonian University, {\L}ojasiewicza 11, 30-348 Krak\'ow, Poland}
\date{\today}

\begin{abstract}
\setlength{\baselineskip}{1.1\baselineskip}
In two-component non-linear Schr\"odinger equations, the force exerted by incident monochromatic plane waves on an embedded dark soliton and on dark-bright-type solitons is investigated, both perturbatively and by numerical simulations.
When the incoming wave is non-vanishing only in the orthogonal component to that of the embedded dark soliton,
its acceleration is in the \textsl{opposite  direction} to that of the incoming wave. This somewhat surprising phenomenon can be attributed to the well known ``negative effective mass'' of the dark soliton.
When a dark-bright soliton, whose effective mass is also negative, is hit by an incoming wave non-vanishing in the component corresponding to the dark soliton, the direction of its acceleration coincides with that of  the incoming wave.
This implies that the net force acting on it is in the \textsl{opposite} direction to that of the incoming wave.
This rather counter-intuitive effect is a yet another manifestation of \textsl{negative radiation pressure} exerted by the incident wave, observed in other systems.
When a dark-bright soliton interacts with an incoming wave in the component of the bright soliton, it accelerates in the opposite direction, hence the force is ``pushing'' it now. We expect that these remarkable effects, in particular the negative radiation pressure, can be experimentally verified in Bose-Einstein condensates.
\end{abstract}

\maketitle
\setlength{\baselineskip}{1.2\baselineskip}

\section{Introduction}
In this paper we consider the interaction of solitons and the sound waves in
a two component, non-linear Schr\"odinger equation (NLSE) in one dimension.
The NLSE is a widely used model, for example in non-linear optics \cite{AgrawalKivshar}, and
in particular it serves to describe Bose-Einstein condensates (BECs) of neutral atoms.
The motivation of our enterprise
is to point out some simple, but somewhat surprising physical phenomena, which are hopefully experimentally observable in BECs.

BECs were realised experimentally for the first time in 1995 \cite{doi:10.1126/science.269.5221.198}, and have been produced in numerous experiments ever since. Many BECs can be described in a mean-field approximation, leading to the NLSE, for which classical field-theoretical methods are appropriate.
Moreover, in a number of situations, the dynamics of solitons in a BEC can be well approximated by restricting the dynamics to one spatial dimension. Often, the trap used in experiments can be approximated by a harmonic potential, and choosing its frequencies in the two chosen dimensions to be much larger than in the remaining third dimension, an effective, quasi-1D cigar-shaped condensate is achieved \cite{Carretero_Gonzalez_2008}. BECs with two distinguishable components are described in a mean-field approximation by coupled nonlinear Schr\"odinger equations (CNLSE) \cite{PhysRevE.91.012924, Deconinck_2003}. Experimentally, such two-component BECs can be achieved either by mixing two different atomic species, e.g., $^{41}$K and $^{87}$Rb \cite{PhysRevLett.89.190404} ($m_1/m_2 \approx 0.47$) or by using two different spin states of the same species \cite{PhysRevLett.81.1539}. Although different kinds of solitons can be found in this system (e.g. \cite{PhysRevLett.109.044102, Degasperis2007}), we focused our considerations on the so-called dark-bright (DB) and dark solitons.

Dark solitons  were studied theoretically since 1971 in the context of BEC \cite{Tsuzuki1971}, and since 1973 in the context of optical fibers \cite{10.1063/1.1654847}. In particular, their dynamics was researched extensively, including multi-soliton interactions \cite{BLOW198555, zakharov1973interaction, Lerner:94}, collisions \cite{Gagnon:93, Thurston:91} and interactions with perturbations \cite{PhysRevE.49.1657, PhysRevE.49.2397, Chen_1998, Huang_1999, PhysRevE.70.066620, PhysRevA.47.1582}. Similarly, dark-bright solitons were a subject of many theoretical articles \cite{PhysRevLett.87.010401, PhysRevE.91.012924, PhysRevA.77.013820, Rajendran_2009, BRAZHNYI2011381, PhysRevA.83.051605}, together with their generalizations to vortex-bright solitons \cite{PhysRevLett.105.160405, PhysRevA.84.053626}, discrete equations \cite{ALVAREZ2011767} and spinor condensates \cite{PhysRevA.77.033612}. Dark and dark-bright solitons were realised experimentally both in BECs \cite{Becker2008, MIDDELKAMP2011642, PhysRevLett.106.065302, PhysRevA.84.053630, PhysRevLett.83.5198, Denschlag2000, doi:10.1126/science.1062527, PhysRevLett.86.2926, bongs2001coherent} and in nonlinear optics \cite{Chen:96, Chen:97, PhysRevLett.77.4011}. A rich review literature on the topic can be found, e.g. \cite{Frantzeskakis_2010, Carretero_Gonzalez_2008, KEVREKIDIS2016140, kevrekidis2008emergent}.

In the present work we show that in the two-component CNLSE, the interaction of dark and dark-bright solitons with incoming small amplitude plane waves can be reasonably well described by standard scattering theory. We derive the force acting on the solitons in terms of scattering data. When the amplitude of the plane waves is sufficiently small, linearization about the soliton provides a tractable approximation with good precision. In this case, the waveform is obtained as a solution of the Bogoliubov-de Gennes equations (BdGe) \cite{PitaevskiiStringari}. The force acting on the soliton can be easily found from momentum conservation. For the case of main interest for us, when an incoming plane wave of amplitude and wave number $a$, $k_1$, is nonzero only in one component, say 1 (dark), then the induced force can be written as:
\begin{equation}
  \label{eq:force}
  F = a^2 \left(
k_1^2 \left(1 + R_1 - T_1 \right)
+ \left(k_2^+ \right)^2 \left(R_2^+- T_2^+ \right)
+ \left(k_2^- \right)^2 \left(R_2^- - T_2^- \right)
\right),
\end{equation}
where $R_{i}$ resp.\ $T_{i}$ denote the transmission resp.\ reflection coefficients
for an incoming wave into channel $i$ and $k_2^\pm$ are the two possible wave numbers in the bright component. In realistic scenarios, for $a$ equal to about 10\% of a DB soliton's dark component amplitude, the accelerations due to this force are up to the order of $10^{-2} \, \mathrm{m/s^2}$ (see Sec. \ref{sec:trap}).

We note that the dynamics due to a force acting on dark resp.\ dark-bright solitons is somewhat counter-intuitive,
since the direction of the force and that of the resulting acceleration points in opposite directions because of their effective \textsl{negative} mass.

The value of the coupling, $g_{12}$, between the two components of the CNLSE plays an important r\^ole,
since for $g_{12}=1$ (in suitable units) the system is integrable \cite{Manakov}, and in this special case the net force exerted by incoming waves on solitons is zero. In fact, for the integrable case, exact solutions corresponding to nonlinear superposition of cnoidal waves and solitons have been constructed \cite{Shin2004}.
We find that quite generally, for $g_{12}\ne1$ in two-component CNLS systems, an incoming plane wave can exert a pulling force on certain solitons  -- referred to as ``tractor beam'' effect, or ``negative radiation pressure'' (NRP).
As it has been already demonstrated for a number of cases in one and two dimensions, in the presence of two scattering channels \textsl{with different dispersion relations}, an incoming plane wave can exert a pulling force on the scatterer \cite{PhysRevD.88.125007}.

The paper is organized as follows. We review briefly some of the main properties of two-component CNLSE (Sec.\ \ref{ssec:GPe}),
exhibit the expressions for the energy and field momentum (Sec.\ \ref{sec:integrals_of_motion}). Next, the linearized equations of motion around a soliton are presented in (Sec.\ \ref{sec:linearization}). Next, we introduce the general notion of the Newtonian approximation using the effective mass and force (Sec.\ \ref{sec:newtonian_motion_and_effective_mass}). After that, we proceed to apply these ideas to the specific cases: dark and dark-bright solitons with a small wave in each component separately (Secs.\ \ref{sec:dark_soliton} and \ref{sec:dark_bright_soliton}). Most importantly, we derive the acceleration of the solitons using an effective model and compare it with numerical simulations of the full CNLSE. Finally, in Sec. \ref{sec:trap} we verify how well the results derived from the homogeneous system apply to a dark-bright soliton in a harmonic trap.

\section{The model}\label{sec:model}
\subsection{Coupled nonlinear Schr\"odinger (Gross--Pitaevskii) equation}\label{ssec:GPe}
In the mean-field regime, a one-dimensional two-component Bose--Einstein condensate can be described by two coupled nonlinear Schr\"odinger equations (CNLSE), also called Gross--Pitaevskii equations, of the form \cite{Yan_2012, PhysRevA.84.053630, Lincoln_D_Carr_2000, app7040388, PhysRevE.91.012924, Deconinck_2003}
\begin{equation}
\label{eq:CNLSE_general}
\begin{split}
i \hbar \partial_t \psi_1 &= - \frac{1}{2 m_1} \partial_{xx} \psi_1 + \left( g_{11} |\psi_1|^2 + g_{12} |\psi_2|^2\right)\psi_1 + V_1(x)\psi_1, \\
i \hbar \partial_t \psi_2 &= - \frac{1}{2 m_2} \partial_{xx} \psi_2 + \left( g_{22} |\psi_2|^2 + g_{12} |\psi_1|^2\right)\psi_2 + V_2(x)\psi_2,
\end{split}
\end{equation}
where $\psi_i$ ($i=1,2$) denote the (complex) wave functions of the two components of the condensate, $m_i$ are their masses and $V_i$ are the trapping potentials experienced by the $i$-th component. If $m_1 = m_2$, the couplings can be written as
$g_{ij} = 2\hbar \omega_\perp a_{ij}$ \cite{Yan_2012, PhysRevA.84.053630, Lincoln_D_Carr_2000, app7040388},
where $a_{ij}$ denote the s-wave scattering lengths between the two components (or within one component in the case of $a_{ii}$) and $\omega_\perp$ is the transverse trapping frequency. The system can be described by the above one-dimensional equations if the frequencies of trapping potentials in the $x$ direction (i.e. $V_i(x)$) are much smaller than $\omega_\perp$. Positive values of $a_{ij}$ (and therefore also of $g_{ij}$) correspond to repulsive interaction between the components $i$ and $j$, whereas a negative value corresponds to the interaction being attractive.

In order to simplify the problem, we reduce the number of parameters used. First, we consider a condensate made
of two different spin states of the same atomic species; therefore, $m_1 = m_2$, and we can rescale $m_i = 1$. Second, we set $g_{ii} = 1$ and keep $g_{12}$ as a free parameter. The former is justified, because the ratio of scattering lengths in experiments is often close to one, e.g., in the mixture of the $\left|2,1\right\rangle$ and $\left|1,-1\right\rangle$ states of $^{87}$Rb without additional tuning it is $a_{11}/a_{12}/a_{22} = 1.03/1/0.97$ \cite{PhysRevLett.81.1539}, or for $\left|1,-1\right\rangle$ and $\left|2,-2\right\rangle$ states: $a_{11}/a_{12}/a_{22} = 1.01/1/1$ \cite{MIDDELKAMP2011642}. Scattering lengths can be manipulated (both their magnitudes and their signs) using Feshbach resonances \cite{RevModPhys.82.1225}. In particular, $a_{12}$ can be tuned independently \cite{condmat5010021, PhysRevLett.100.210402, PhysRevA.82.033609}, and therefore different values of the $g_{12}$ coefficient are achievable in experiments. Finally, we assume that $V_i = 0$, which should be a valid first approximation: we will verify that later in Sec. \ref{sec:trap}. Therefore, we assume $a_{11} = a_{22}$ and take $m = m_1 = m_2$ as the unit of mass, $\hbar/(2 a_{11} m \omega_\perp )$ as the length unit and $\hbar/(4 a_{11}^2 m \omega_\perp^2)$ as the unit of time. Then, the set of equations (\ref{eq:CNLSE_general}) takes the form
\begin{equation}
\label{eq:CNLSE}
\begin{split}
i \partial_t \psi_1 &= - \frac{1}{2} \partial_{xx} \psi_1 + \left( |\psi_1|^2 + g_{12} |\psi_2|^2\right)\psi_1\,, \\
i \partial_t \psi_2 &= - \frac{1}{2} \partial_{xx} \psi_2 + \left( |\psi_2|^2 + g_{12} |\psi_1|^2\right)\psi_2\,.
\end{split}
\end{equation}
The system of equations \eqref{eq:CNLSE} has various type of solitonic solutions, see the review \cite{Frantzeskakis_2010}. The simplest solutions correspond to the embedding of a scalar soliton into one of the components.
We shall consider embedded dark solitons, for which the probability density has a dip, and dark-bright (DB) solitons, having a dip and a peak of the probability density in one, resp.\ in the other component of $(\psi_1\,,\psi_2)$. When $g_{12} = 1$, Eqs.\ (\ref{eq:CNLSE}) correspond to the Manakov system \cite{Manakov} which is known to be integrable, c.f. also Ref.\ \cite{Wang2010}. For the integrable case, the DB solitons are known analytically, while for values of $g_{12}\ne1$ we have solved Eqs.\ \eqref{eq:CNLSE}  numerically. For such values of $g_{12}\ne1$, we have shown analytically and also confirmed by numerical simulations that incoming sound waves (referred to as ``radiation'') do exert a force on the solitons.

Usually, particle-like objects such as solitons are pushed in the direction of propagation of an incoming wave. However, in some cases the wave pulls a soliton in the \textsl{opposite} direction. We refer to such a situation as \textsl{negative radiation pressure} (NRP). In the case of dark and dark-bright solitons, this definition needs an additional clarification. Such solitons have negative effective mass, therefore their acceleration has an opposite sign to the effective force. Thus, we define a positive/negative radiation pressure (PRP/NRP) as a setup in which the \textsl{force} has the same/opposite sign to the direction of the incoming wave. This means that, in terms of an acceleration of dark and DB solitons, PRP and NRP correspond to a wave pulling and pushing a soliton respectively. We remark that in the literature on NRP \cite{Romanczukiewicz:2003tn, Romanczukiewicz:2008hi, Forgacs:2008az, PhysRevLett.107.091602, ROMANCZUKIEWICZ2017295, PhysRevD.88.125007}, in the systems considered up to now only positive masses occured; therefore the NRP exerted by incident plane waves has manifested itself by a pulling effect.

\subsection{Integrals of motion}
\label{sec:integrals_of_motion}
The Lagrangian density corresponding to the CNLSE (\ref{eq:CNLSE}) is \cite{PhysRevA.82.053601}:
\begin{equation}
\label{eq:CNLSE_lagrangian}
\mathcal{L} = \frac{1}{2} \sum_{i=1,2} \left[ i \left( \psi_i^* \partial_t \psi_i - \psi_i \partial_t \psi_i^* \right) - |\partial_x \psi_i|^2 - |\psi_i|^4 - g_{12} |\psi_1|^2 |\psi_2|^2 \right]\,.
\end{equation}
Using the symmetries of the Lagrangian (\ref{eq:CNLSE_lagrangian}) and the Noether theorem (details in the Appendix \ref{sec:integrals_of_motion_derivation}), the total energy and momentum are derived as
\begin{align}
\label{eq:total_energy_explicit}
E &=  \frac12 \sum_{i=1,2} \int_{-\infty}^{\infty} \left( |\partial_x \psi_i|^2 + |\psi_i|^4 + g_{12} |\psi_1|^2 |\psi_2|^2  \right) {\rm d}x\,, \\
\label{eq:total_momentum_explicit}
P &= \frac{i}{2} \sum_{i=1,2} \int_{-\infty}^{\infty} \left(\psi_i \partial_x \psi_i^*  - \psi_i^* \partial_x \psi_i\right) {\rm d}x\,,
\end{align}
and it is shown that they obey the equations
\begin{align}
\label{eq:energy_momentum_conservation_explicit_E_total}
\partial_t E &= \frac12 \sum_{i=1,2} \left. \left( \partial_x \psi_i^* \partial_t \psi_i + \partial_x \psi_i \partial_t \psi_i^* \right) \right|_{-\infty}^{\infty}\,, \\
\label{eq:energy_momentum_conservation_explicit_P_total}
\partial_t P &= \frac12 \sum_{i=1,2} \left. \left(- i \left( \psi_i^* \partial_t \psi_i - \psi_i \partial_t \psi_i^* \right) - |\partial_x \psi_i|^2 + |\psi_i|^4 + g_{12} |\psi_1|^2 |\psi_2|^2 \right) \right|_{-\infty}^{\infty}\,.
\end{align}
It is also worth noting that, due to the $U(1) \times U(1)$ symmetry, the solutions of Eq.\ (\ref{eq:CNLSE}) obey the following continuity equations (even if we include the trapping potential):
\begin{equation}
\label{eq:continuity}
\partial_t |\psi_i|^2 + \partial_x J_i = 0\,,
\end{equation}
where
\begin{equation}
\label{eq:probability_current}
J_i = \frac{i}{2} \left(\psi_i \partial_x \psi_i^*  - \psi_i^* \partial_x \psi_i\right)\,.
\end{equation}
Note that $P = \sum_{i=1,2} \int_{-\infty}^{\infty} J_i \, {\rm d}x$. Integrating Eq.\ (\ref{eq:continuity}) over the whole space and using the Newton-Leibniz theorem, we obtain
\begin{equation}
\partial_t \int_{-\infty}^{\infty} |\psi_i|^2 {\rm d}x = - \left. J_i \right|_{-\infty}^{\infty}\,.
\end{equation}
We can choose the normalization as $\int |\psi_i(x,t)|^2 {\rm d}x = N_i$, where $N_i$ is the number of atoms in the $i$-th component. These particle numbers are conserved.

\subsection{Linearization around a soliton}
\label{sec:linearization}
We shall consider stationary solitons of Eqs.\ (\ref{eq:CNLSE}) of the form $\psi_i(x,t) = e^{- i \mu_i t} \Phi_i(x)$, where $\Phi_i(x)$ are real functions satisfying the following equations:
\begin{equation}
\label{eq:CNLSE_stationary}
\begin{split}
\mu_1 \Phi_1 &= - \frac{1}{2} \partial_{xx} \Phi_1 + \left( |\Phi_1|^2 + g_{12} |\Phi_2|^2\right)\Phi_1\,, \\
\mu_2 \Phi_2 &= - \frac{1}{2} \partial_{xx} \Phi_2 + \left( |\Phi_2|^2 + g_{12} |\Phi_1|^2\right)\Phi_2\,.
\end{split}
\end{equation}
If the wave function is normalized to the number of atoms in each component, then $\mu_i$ are determined from these normalization conditions, and they are interpreted as chemical potentials \cite{Trippenbach_2000}.

Let us consider a small perturbation of a soliton solution of Eqs.\ \eqref{eq:CNLSE}, of the form:
\begin{equation}
\label{eq:soliton_perturbed}
\psi_i(x,t) = e^{- i \mu_i t} \left( \Phi_i(x) + a \xi_i(x,t) \right),
\end{equation}
where the parameter of the perturbation, $a\ll 1$. Moreover, let us make an ansatz
\begin{equation}
\xi_i(x,t) = e^{i \tilde{\omega} t} \xi_i^+(x) + e^{-i \tilde{\omega} t} \xi_i^-(x).
\end{equation}
After inserting these Ans\"atze into Eq.\ (\ref{eq:CNLSE}) and keeping only terms linear in $a$, we obtain that $\xi_i^-$ and $\xi_i^+$ satisfy:
\begin{equation}
\label{eq:linearized}
\left(-\frac{1}{2} \partial_{xx} + \mathbf{M}\right) \Xi = \mathbf{diag}(\mu_1 + \tilde{\omega}, \mu_1 - \tilde{\omega}, \mu_2 + \tilde{\omega}, \mu_2 - \tilde{\omega}) \Xi,
\end{equation}
where $\mathbf{diag}$ means diagonal matrix, $\Xi$ stands for the vector
\begin{equation}
\Xi = (\xi_1^-, {\xi_1^+}^*, \xi_2^-, {\xi_2^+}^*)^\mathrm{T}
\end{equation}
and
\begin{equation}
\label{eq:matrix_M}
\mathbf{M} = \left( \begin{matrix}
2 \Phi_1^2 + g_{12} \Phi_2^2  & \Phi_1^2 & g_{12} \Phi_1 \Phi_2 & g_{12} \Phi_1 \Phi_2 \\
\Phi_1^2 & 2 \Phi_1^2 + g_{12} \Phi_2^2 & g_{12} \Phi_1 \Phi_2 &  g_{12} \Phi_1 \Phi_2 \\
g_{12} \Phi_1 \Phi_2 & g_{12} \Phi_1 \Phi_2 & g_{12} \Phi_1^2 + 2 \Phi_2^2 & \Phi_2^2 \\
g_{12} \Phi_1 \Phi_2 & g_{12} \Phi_1 \Phi_2 & \Phi_2^2 & g_{12} \Phi_1^2 + 2 \Phi_2^2
\end{matrix} \right).
\end{equation}
Eq.\ (\ref{eq:linearized}) is the generalization of the BdGe \cite{PitaevskiiStringari} to the two-component CNLSE. Here we shall consider plane wave solutions of Eq.\ (\ref{eq:linearized}), but also note, that the existence of bound state solutions with complex $\tilde\omega$ would signal instability of the soliton. Such stability analyses have been performed, e.g., in Refs.\ \cite{PhysRevE.91.012924, KEVREKIDIS2016140} in similar coupled models, or in Ref.\ \cite{tran1992stability, walczak2011exact, pelinovsky2022stability} for a single component NLSE. In our case, the numerical technique used for finding the soliton (gradient descent) already ensures that unstable solitons are not found.

We shall consider a setup consisting of a soliton and a wave. The wave is incoming from $- \infty$ in one of the two components of BEC and is moving to the right. It corresponds to equation (\ref{eq:soliton_perturbed}) with $a$ interpreted as the amplitude (and the appropriate boundary conditions discussed later). Using the linearization described above, the wave can be written as
\begin{equation}
\label{eq:linearized_split}
 a e^{- i \mu_i t} \xi_i(x, t) =  a e^{- i (\mu_i - \tilde{\omega}) t}  \xi_i^+(x) + a e^{- i (\mu_i + \tilde{\omega}) t} \xi_i^-(x)
\end{equation}
and we can denote its frequencies as
\begin{equation}
\label{eq:linearized_frequencies}
\omega_i^{\pm} = \mu_i \mp \tilde{\omega},
\end{equation}
which correspond to these in the r.h.s. of Eq.\ (\ref{eq:linearized}). Sometimes we shall loosely refer to $\tilde{\omega}$ as the frequency, but the true frequencies of the wave are given by $\omega_i^{\pm}$. We assume that asymptotically these waves have a form of monochromatic plane waves. We shall define transmission and reflection coefficients, separately for each of the examples, as coefficients of the asymptotic plane wave modes in a solution.

\section{Newtonian motion and the effective mass of solitons}
\label{sec:newtonian_motion_and_effective_mass}
We expect both the dark and the dark-bright solitons of the NLSE to behave as Newtonian particles in the first approximation, albeit with unusual dynamics, due to their negative effective masses. In this present context, we refer to Ref.\ \cite{Busch-Anglin2000} on the motion of dark solitons, and for a recent review on the dynamics of solitons in the vector NLSE see Ref.\ \cite{KEVREKIDIS2016140}.

More precisely, we shall assume that in the presence of perturbations, the solitons do not change their shape, and that we can treat the centre of the soliton $x_0$ solely as a function of time, reducing the problem to one-dimensional classical dynamics. In this description, $x_0(t)$ is expected to obey the equation
\begin{equation}
M \ddot{x}_0(t) = F\,,
\end{equation}
where $M$, resp.\ $F$ is the effective mass resp.\ force. These quantities are obtained from the integrals of motion discussed in Section \ref{sec:integrals_of_motion}.
The description of the dynamics of the dark soliton is complicated by the fact, that the wavefunction describes the soliton on the top of a constant background \cite{Busch-Anglin2000}. 

The energy, $E$, and the momentum $P$ of the dark soliton have to be defined carefully, they have to be ``renormalized'' in order to subtract the contribution from the background \cite{PhysRevE.49.1657, Frantzeskakis_2010}. The renormalized quantities $P_s$ and $E_s$ will be given separately for the dark and dark-bright soliton. See Secs.\ \ref{sec:dark_soliton} and \ref{sec:dark_bright_soliton}.

A useful definition of the effective mass, $M$, from the ``renormalized'' total energy, $E_s$, resp.\ momentum, $P_s$, for a soliton moving with velocity $v$, is given as:
\begin{equation}
M = \left. \frac{{\rm d}^2 E_s}{{\rm d}v^2}\right|_{v=0} = \left. \frac{{\rm d} P_s}{{\rm d}v}\right|_{v=0}\,.
\end{equation}
If the soliton is indeed moving according to Newton's law, $M$ computed from $E_s$ should match that derived from $P_s$.

The effective force exerted by the sound waves on the soliton can be obtained as the time derivative of the total momentum, $\partial_t P$. Since the ``renormalization corrections'' are time-independent, $\partial_t P=\partial_t P_s$.
We shall be interested in the force averaged over a period of the incoming wave, thus we are led to define the effective force as:
\begin{equation}
F = \langle \partial_t P \rangle_T\,,
\end{equation}
where $P$ is the total momentum including the radiation and $\langle \cdot \rangle_T$ is the average over a period.
We note that to evaluate Eq.\ (\ref{eq:energy_momentum_conservation_explicit_P_total}) it is sufficient to know the asymptotic form of the radiation in order to compute the effective force.
In the computation of $F$, we can only keep the contributions of order $a^2$ since we have the solution up to linear order in the amplitude. In this linearized approximation (which turns to be quite efficient), one can easily obtain the results for any incoming wave-form.

\section{Dark soliton}\label{sec:dark_soliton}
\subsection{The dark soliton solution}
\label{sec:dark_soliton_solution}
A particular solution of Eq.\ (\ref{eq:CNLSE}) with arbitrary $g_{12}$ is a (scalar) dark soliton centered at $x_0$ \cite{PhysRevE.91.012924}, embedded into the vector NLSE:
\begin{equation}\label{darksoliton}
\begin{split}
\psi_1 &=  e^{-i \mu t} \sqrt{\mu} \tanh (\sqrt{\mu}  (x-x_0))\,, \quad
\psi_2 =  0\,,
\end{split}
\end{equation}
with chemical potential, $\mu_1 = \mu>0$. Such a soliton can be understood as a dip in the probability density obtained from the collective wave function of atoms in the condensate. We will consider two scenarios: in the first case the additional atoms will be present in the second component in the form of a plane wave (a system still described by the two-component model), and in the second case the second component will be completely absent (a problem reduced to the one-component NLSE).

In order to examine the effective force exerted by an incoming plane wave on a dark soliton, we need to analyze the linearized equations. The relations between $\tilde{\omega}$ (see section \ref{sec:linearization}) and the wave numbers, $k_i$, is obtained from the $x\to\pm\infty$ asymptotic form of equation (\ref{eq:linearized}). Knowing that $\Phi_1(x) \to \pm \sqrt{\mu}$ and $\Phi_1''(x) \to 0$ as $x \to \pm \infty$, the asymptotic form of the matrix from Eq.\ (\ref{eq:matrix_M}) is
\begin{equation}
\mathbf{M}_{x \to \pm \infty} = \mu \left( \begin{matrix}
2 & 1 & 0 & 0 \\
1 & 2 & 0 &  0 \\
0 & 0 & g_{12} & 0 \\
0 & 0 & 0 & g_{12}
\end{matrix} \right)\,.
\end{equation}
The linearized equations (\ref{eq:linearized}) with this $\mathbf{M}$ can easily be diagonalized and solved, obtaining:
\begin{equation}
\label{eq:scalar_asymptotic_linearized_1}
\begin{split}
\xi_1^-(x) &= A e^{i k_1 x} + B e^{-i k_1 x}\,, \\
\xi_1^+(x) &= \left(-\frac{\frac{k_1^2}{2}-\tilde{\omega}}{\mu} -1 \right)
\left(A^* e^{-i k_1 x} + B^* e^{i k_1 x}\right)\,,
\end{split}
\end{equation}
 where $A$ and $B$ are arbitrary constants, and
\begin{equation}
k_1 = \sqrt{2} \sqrt{\sqrt{\mu^2 + \tilde{\omega}^2} - \mu}\,.
\end{equation}
In the above solution we omitted the solutions with imaginary wavenumber, since they describe non-propagating solutions, and therefore do not carry a momentum. Note that for $\tilde{\omega} \neq 0$ the wavenumber, $k_1$, is real thus Eqs.\ \eqref{eq:scalar_asymptotic_linearized_1} describe propagating (travelling) waves.

As mentioned earlier, we shall consider waves coming from the left and moving to the right. The direction of the travelling wave is determined by a relative sign of the wave frequency and its wavevector. Assuming an incoming wave with $e^{i k_1 x}$ for $\omega_1^+$ and (to be consistent with the above solution) $e^{-i k_1 x}$ for $\omega_1^-$, the condition for moving to the right is that the frequency $\omega_1^\pm = \mu \mp \tilde{\omega}$ is positive/negative respectively (cf.\ Eqs.\  (\ref{eq:linearized_split}) and (\ref{eq:linearized_frequencies})). This can be transformed into the following conditions for $\tilde{\omega}$: $\tilde{\omega} < \mu$ for $\omega_1^+$ and $\tilde{\omega} < - \mu$ for $\omega_1^-$. Therefore, for $\tilde{\omega} < - \mu$ both waves propagate to the right.

In the second component, the equations are already diagonal. Note that in this case $\mu_2$ is arbitrary and choosing it can be interpreted as fixing a reference point for $\tilde{\omega}$. For simplicity, let us put $\mu_2 = 0$, then we can interpret $\mp \tilde{\omega} = \omega_2^\pm$ simply as the frequency of the incoming wave in the second sector. Then the solutions are
\begin{equation}
\label{eq:scalar_asymptotic_linearized_2}
\begin{split}
\xi_2^-(x) &= C e^{i k_2^- x} + D e^{-i k_2^- x}\,, \\
\xi_2^+(x) &= E e^{i k_2^+ x} + F e^{-i k_2^+ x}\,,
\end{split}
\end{equation}
where
\begin{equation}
k_2^\pm = \sqrt{2} \sqrt{\mp \tilde{\omega} - \mu g_{12}}\,,
\end{equation}
and $C, D, E, F$ are arbitrary constants. These waves propagate when $k_2^\pm$ is real (and nonzero), that is, when $\mp \tilde{\omega} > \mu g_{12}$. This means that for $g_{12} < 0$ there exists a range of $\tilde{\omega}$ in which both waves can propagate with the same $\tilde{\omega}$ and for $g_{12} \geq 0$ with fixed $\tilde{\omega}$ only one (or neither) of the waves can propagate.

Let us analyse how the parameter $\tilde{\omega}$ affects the direction of propagation of the waves in the second component. Using the same logic as for the first component (assuming an incoming wave with $e^{i k_2^\pm x}$), we get the conditions for the range of $\tilde{\omega}$ in which the waves in the second component are moving to the right. Namely, we obtain: $\tilde{\omega} < 0$ for $\omega_2^+$ and $\tilde{\omega} > 0$ for $\omega_2^-$. This means that only one of them moves to the right for a given $\tilde{\omega}$.

In order to find the effective mass of the soliton, we repeat the derivation of the renormalized momentum done in \cite{PhysRevE.49.1657} (for additional details see also \cite{BARASHENKOV1993114}) and obtain the effective mass from it following \cite{Frantzeskakis_2010}. First, we consider the total momentum, $P_s$, of a scalar dark soliton moving with a constant velocity $v$  (often called the gray soliton):
\begin{equation}
\label{eq:moving_scalar_dark}
\begin{split}
\psi_1 &=  e^{-i \mu t} \left(i v +  \sqrt{\mu - v^2} \tanh (\sqrt{\mu - v^2}  (x-x_0-vt)) \right)\,, \\
\psi_2 &=  0\,.
\end{split}
\end{equation}
Note, that the moving soliton becomes shallower and shallower as the velocity increases, finally vanishing when $v^2 = \mu$, which defines its maximal velocity. However, the above wavefunction describes a dark soliton on top of a background, and we are interested in the total momentum of the soliton. Let us note that the solution with constant probability density, corresponding to the asymptotics of (\ref{eq:moving_scalar_dark}) (i.e. $|\psi_1|^2 = \mu$ and $|\psi_2|^2 = 0$) is of the form
\begin{equation}
\begin{split}
\psi_1 &=  \sqrt{\mu} e^{-i (\mu + q^2/2) t} e^{i q x}\,, \\
\psi_2 &=  0\,,
\end{split}
\end{equation}
with some real $q$. Since we are interested in a non-moving background, we choose $q=0$, yielding the background part of the dark soliton. However, we have to take into account the phase change induced by the presence of the soliton (see \cite{PhysRevE.49.1657}). Therefore, we assume that the background (in the first component) has the form:
\begin{equation}
\label{eq:moving_scalar_dark_background}
\psi_b =  \sqrt{\mu} e^{-i \mu t}  e^{i k(x) x}\,,
\end{equation}
where $k$ is some real function of $x$, reflecting the phase change induced by a soliton. It will turn out that the explicit form of $k(x)$ is not needed. Although the probability density of the background  is constant (equal to $\mu$), the total momentum contribution also depends on the phase. Inserting Eq.\ (\ref{eq:moving_scalar_dark_background}) to Eq.\  (\ref{eq:total_momentum_explicit}) one obtains that the contribution of the background can be written as:
\begin{equation}
\label{eq:scalar_dark_momentum_background_contribution}
\mu \Delta \phi \equiv \mu \int_{-\infty}^{\infty} \left( k(x) + x k'(x) \right){\rm d}x = \mu x k(x)_{-\infty}^{\infty}\,.
\end{equation}
Comparing with Eq.\ (\ref{eq:moving_scalar_dark_background}), $\Delta \phi$ is readily identified with the induced phase change of the background. Therefore, the total momentum of the `pure soliton' can be expressed as (cf. \cite{PhysRevE.49.1657, Frantzeskakis_2010})
\begin{equation}\label{eq:dark_momentum}
P_s = \frac{i}{2} \int_{-\infty}^{\infty} \left(\psi_1 \partial_x \psi_1^*  - \psi_1^* \partial_x \psi_1 \right) {\rm d}x - \mu \Delta \phi\,.
\end{equation}
The phase change, $\Delta \phi$, can be easily computed from the asymptotics of Eq.\ (\ref{eq:moving_scalar_dark}):
\begin{equation}
\Delta \phi = - 2 \arctan \left( \frac{\sqrt{\mu - v^2}}{v} \right)\,.
\end{equation}
Using this, we obtain from Eq.\ \eqref{eq:dark_momentum} the momentum corresponding to the soliton:
\begin{equation}
\label{eq:scalar_dark_momentum}
P_s = - 2 v \sqrt{\mu - v^2} + 2 \mu \arctan \left( \frac{\sqrt{\mu - v^2}}{v} \right)\,,
\end{equation}
which allows us to compute its effective mass as  (cf. \cite{Frantzeskakis_2010})
\begin{equation}
\label{eq:scalar_dark_mass}
M = \left. \frac{d P_s}{d v}\right|_{v \to 0} = -4 \sqrt{\mu}\,.
\end{equation}
Intuitively, the negative sign is not a surprise, because a dark soliton is, as mentioned before, a dip in the collective probability density of atoms. The same mass is obtained using the renormalized energy \cite{PhysRevE.49.1657} (see section \ref{sec:newtonian_motion_and_effective_mass})
\begin{equation}
E_s = \frac12 \int_{-\infty}^{\infty} \left( |\partial_x \psi_1|^2 + \left(|\psi_1|^2 - \mu \right)^2 \right) {\rm d}x\,.
\end{equation}

\subsection{Wave in the second component}\label{ssec:DS_wave_2nd_component}

Let us now consider the interaction of an embedded dark soliton in the 1st component and an incoming wave in the second one.
In this case one can obtain the analytic solutions of the linearized equations for the waveform in the second component. Since in the case of an embedded  dark soliton, \eqref{darksoliton}, the linearized equations \eqref{eq:linearized} for $\xi_1$ and $\xi_2$ are decoupled from each other, we may simply put $\xi_1 = 0$. Then Eqs.\ (\ref{eq:linearized}) are reduced to:
\begin{equation}
\label{eq:scalar_dark_linearized_equations_2}
\begin{split}
- \frac12 \partial_{xx} \xi_2^-(x) + g_{12} \mu \tanh^2(\sqrt{\mu} x) \xi_2^-(x) &= \tilde{\omega} \xi_2^-(x)\,, \\
- \frac12 \partial_{xx} \xi_2^+(x) + g_{12} \mu \tanh^2(\sqrt{\mu} x) \xi_2^+(x) &= - \tilde{\omega} \xi_2^+(x)\,.
\end{split}
\end{equation}
Regular solutions of Eq.\ \eqref{eq:scalar_dark_linearized_equations_2} can be expressed in terms of associated Legendre functions of the first kind:
\begin{equation}
\label{eq:scalar_dark_linearized_solution_2}
\xi_2^\pm(x) = A P_{\lambda}^{ik_2^\pm/\sqrt{\mu}}\left(\tanh(\sqrt{\mu} x) \right)\,,
\end{equation}
where $\lambda = \frac12 \left( \sqrt{1 + 8 g_{12}} - 1 \right)$ and $A$ is a normalization factor.

Since our boundary conditions correspond to a wave coming from $x=-\infty$, we impose the following asymptotic behaviour on $\xi_2$:
\begin{equation}
\begin{split}
\xi_2^\pm (x) &\xrightarrow[x \to -\infty]{} e^{i k_2^\pm x} + r_2^\pm e^{-i k_2^\pm x}\,,   \\
\xi_2^\pm (x) &\xrightarrow[x \to +\infty]{} t_2^\pm e^{i k_2^\pm x}\,.
\end{split}
\end{equation}
This asymptotics can be ensured by choosing the normalization constant, $A$, in Eq.\ (\ref{eq:scalar_dark_linearized_solution_2}) appropriately.
The reflection resp.\ transmission coefficients are defined as $R_2^\pm = \left|r_2^\pm \right|^2$ resp.\ $T_2^\pm = \left|t_2^\pm \right|^2$.
The reflection resp.\ transmission coefficients can be written as:
\begin{equation}
\label{eq:scalar_second_component_R_T}
\begin{split}
R_2^\pm &= \frac{2 \sin^2(\pi \lambda)}{\cosh \left(\frac{2 \pi k_2^\pm}{\sqrt{\mu}}\right) - \cos (2 \pi \lambda )}\,, \\
T_2^\pm &= \frac{2 \sinh^2\left(\frac{\pi k_2^\pm}{\sqrt{\mu}}\right)}{\cosh \left(\frac{2 \pi k_2^\pm}{\sqrt{\mu}}\right)-\cos (2 \pi \lambda )}\,.
\end{split}
\end{equation}
One can check that the following relation is satisfied:
\begin{equation}
\label{eq:R_T_1_scalar}
R_2^\pm + T_2^\pm = 1\,.
\end{equation}
Note that the reflection coefficient is zero not only for $g_{12} = 1$, but for any value of $g_{12}$ such that $\lambda$ is an integer.

We now investigate the dynamics of a dark soliton embedded to the first component under the influence of an incident plane wave coming from $x=-\infty$ embedded to the second component. We shall stick to the linearized approximation and we assume the amplitude of the wave, $a$, to be sufficiently small.
Let us consider the setup in which $\tilde{\omega}$ is such that only one of $\xi_2^\pm$ waves has a real wavenumber. Then, we can omit the other one ($\xi_2^+$ or $\xi_2^-$), since it describes a non-propagating solution and does not carry a momentum. In terms of full wavefunctions $\psi_i$ this setup has the following asymptotics:
\begin{equation}
\begin{split}
\psi_1(x,t) &\xrightarrow[x \to -\infty]{} - \sqrt{\mu } e^{-i \mu  t}\,,   \\
\psi_1(x,t) &\xrightarrow[x \to +\infty]{} \sqrt{\mu } e^{-i \mu  t}\,, \\
\psi_2(x,t) &\xrightarrow[x \to -\infty]{} a e^{- i \omega_2^\pm t}  (e^{i k_2^\pm x} + r_2^\pm e^{-i k_2^\pm x})\,,   \\
\psi_2(x,t) &\xrightarrow[x \to +\infty]{} a e^{- i \omega_2^\pm t}  t_2^\pm e^{i k_2^\pm x}\,.
\end{split}
\end{equation}
In order to approximate the acceleration of the soliton, we shall assume Newtonian motion, with the force stemming from the radiation pressure, averaged over a period of the incoming wave.

The force is derived from Eq.\ (\ref{eq:energy_momentum_conservation_explicit_P_total}). Firstly, we substitute the above asymptotic form into this equation. Then we average it over time for the period $T = 2\pi/\tilde{\omega}$ (which in this case does not change anything) and omit the terms of the order higher than $a^2$ (since $a$ is small). Finally, we substitute $\tilde{\omega}$ with the appropriate dispersion relation with $k_2^\pm$, derived in the previous subsection, obtaining the force:
\begin{equation}
\label{eq:scalar_second_component_force}
F = \langle \partial_t P \rangle_T = a^2 \left(k_2^\pm \right)^2 \left(1 + R_2^\pm - T_2^\pm \right)\,,
\end{equation}
where $P$ is the total momentum and $\langle \cdot \rangle_T$ means the average over the period. Using the relation (\ref{eq:R_T_1_scalar}) the force can be simplified to $F  = 2 a^2 \left(k_2^\pm \right)^2 R_2^\pm$, therefore in this case reflectionlessness implies no force (of the assumed order $a^2$).

Let us briefly consider the range of $\tilde{\omega}$ for which $\xi_2^+$ and $\xi_2^-$ are both travelling waves. Then, the initial wave $\xi_2^\pm$ can scatter into both $\xi_2^+$ and $\xi_2^-$, and the analogous derivation leads to the force
\begin{equation}
\label{eq:scalar_second_component_force_both}
F = a^2 \left( \left(k_2^\pm\right)^2 + \left(k_2^+\right)^2 \left(\tilde{R}_2^+ - \tilde{T}_2^+ \right) + \left(k_2^- \right)^2 \left(\tilde{R}_2^- - \tilde{T}_2^-\right) \right)\,,
\end{equation}
where the incoming wave has a wavenumber $k_2^\pm$. However, after considering the boundary conditions, $\tilde{R}_2^\pm$ and $\tilde{T}_2^\pm$ are equal to $R_2^\pm$, $T_2^\pm$ from Eq.\ (\ref{eq:scalar_second_component_R_T}) only for the incoming wave, and for the other one are equal to zero. Therefore, the above expression ultimately reduces to Eq.\ (\ref{eq:scalar_second_component_force}). If both are the incoming waves, only one of them is moving to the right for given $\tilde{\omega}$ (see the discussion in \ref{sec:dark_soliton_solution}) and also the effective force is simply a sum of individual forces, so it is not interesting at this point.

Finally, using the reflection coefficient given by Eq.\ (\ref{eq:scalar_second_component_R_T}), mass derived in the previous section and the above $F$, we can derive the explicit form of acceleration exerted on the scalar dark soliton in the first component by a wave with frequency $\tilde{\omega}$ in the second component. For the incoming wave with a wavenumber $k_2^\pm$ it is
\begin{equation}
\label{eq:scalar_second_component_acceleration}
\ddot{x}_0 = - \frac{a^2}{\sqrt{\mu}}
\frac{\left(k_2^\pm\right)^2 \sin^2(\pi \lambda)}{\cosh \left(\frac{2 \pi k_2^\pm}{\sqrt{\mu}}\right) - \cos (2 \pi \lambda)}\,,
\end{equation}
where $x_0$ is interpreted as the position of the  the soliton. Note that although the force (\ref{eq:scalar_second_component_force}) is always nonnegative, the acceleration is always nonpositive due to the negative effective mass (\ref{eq:scalar_dark_mass}), therefore we observe either the positive radiation pressure (i.e., positive force) or no pressure at all.

\begin{figure}
	\includegraphics[width=\columnwidth]{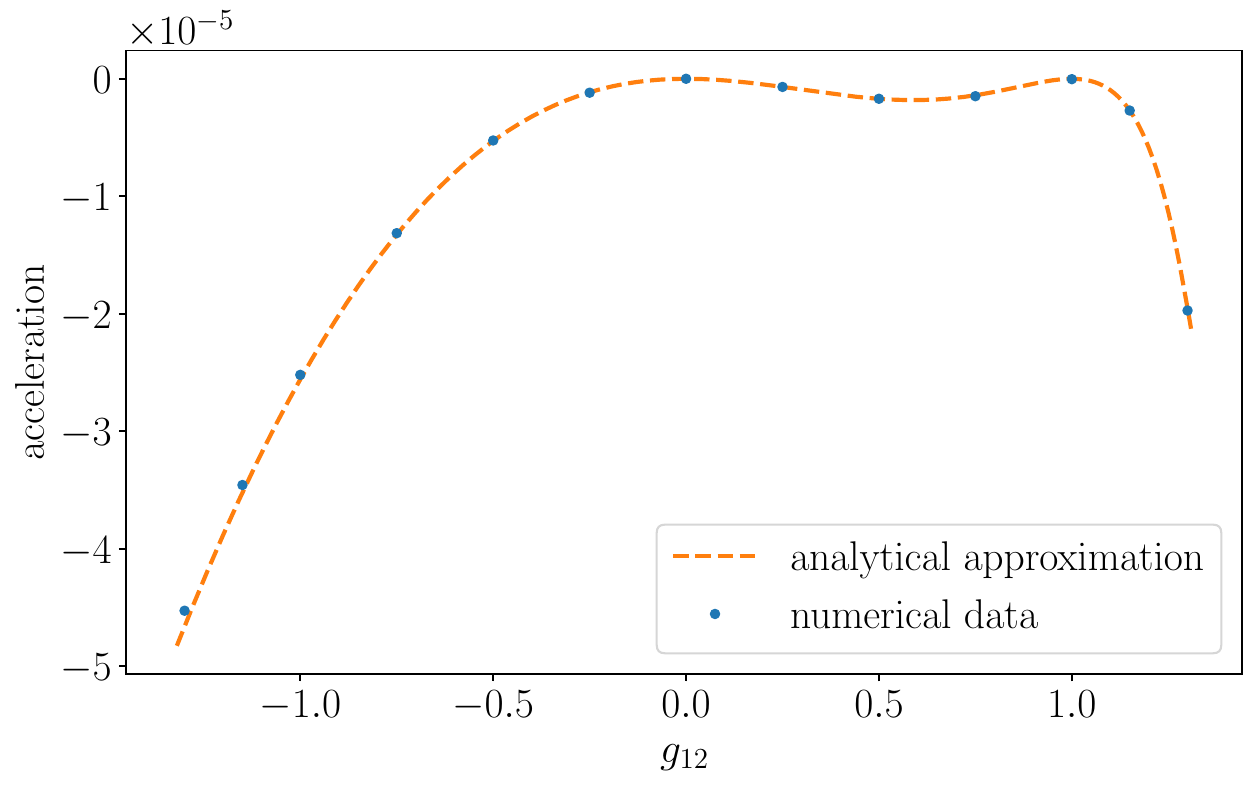}
	\caption{Acceleration of a dark soliton with $\mu = 1$ under the influence of the wave in the second component coming from the left with a frequency $\omega_2^+ = -\tilde{\omega}= 1.4$ and amplitude $a = 0.05$ for different values of $g_{12}$.}
	\label{fig:scalar_g_2}
\end{figure}

\begin{figure}
	\includegraphics[width=\columnwidth]{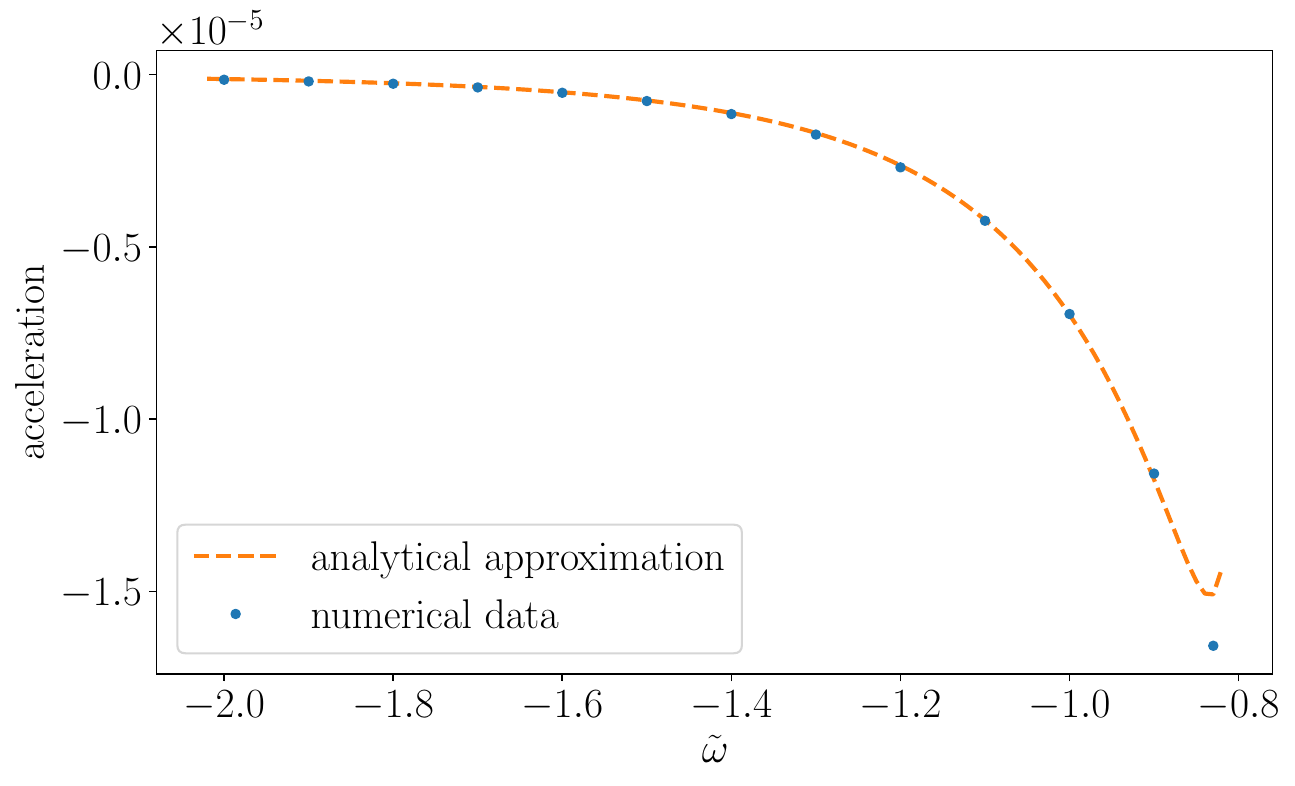}
	\caption{Acceleration of a dark soliton with $\mu = 1$ under the influence of the wave in the second component coming from the left with different frequencies $\omega_2^+ = -\tilde{\omega}$, $g_{12} = 0.8$ and amplitude $a = 0.05$.}
	\label{fig:scalar_frequency_2}
\end{figure}

\begin{figure}
	\includegraphics[width=\columnwidth]{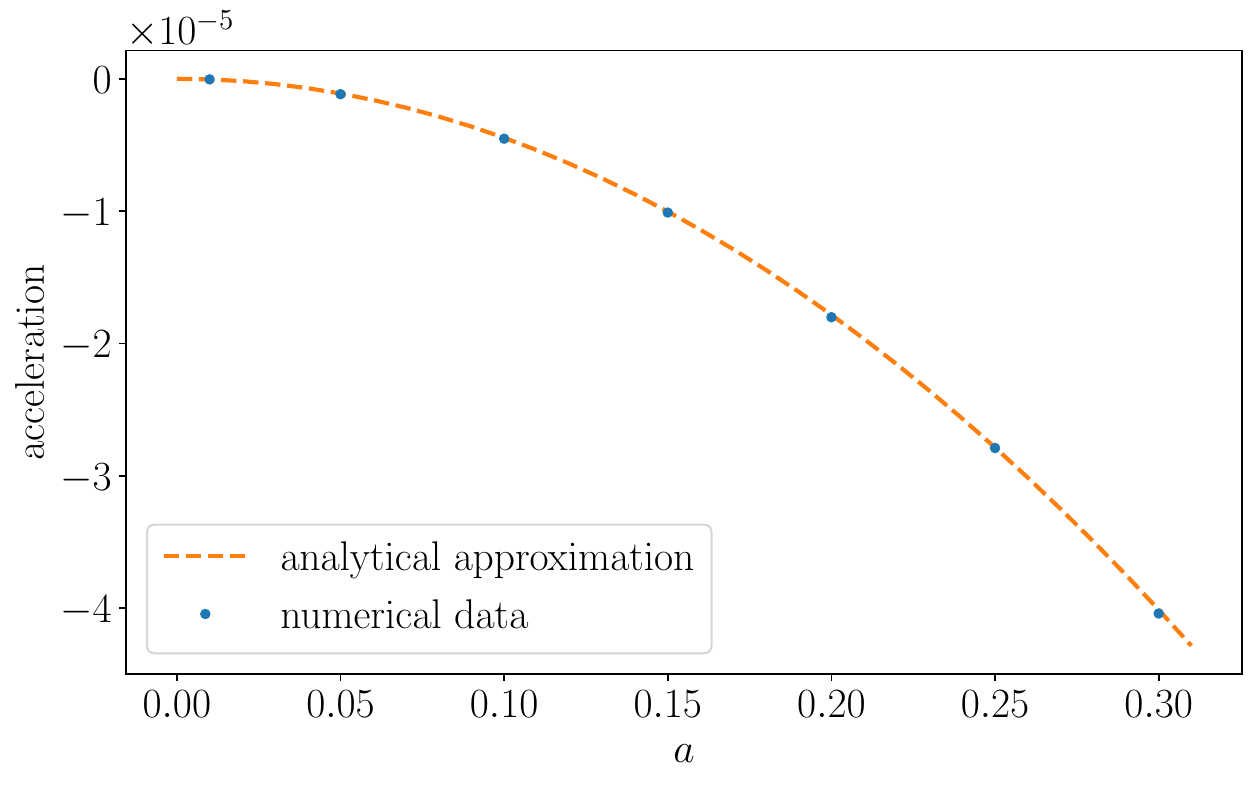}
	\caption{Acceleration of a dark soliton with $\mu = 1$ under the influence of the wave in the second component coming from the left with a frequency $\omega_2^+ = -\tilde{\omega}= 1.4$, $g_{12} = 0.8$ and different amplitudes.}
	\label{fig:scalar_amplitude_2}
\end{figure}

\begin{figure}
	\includegraphics[width=\columnwidth]{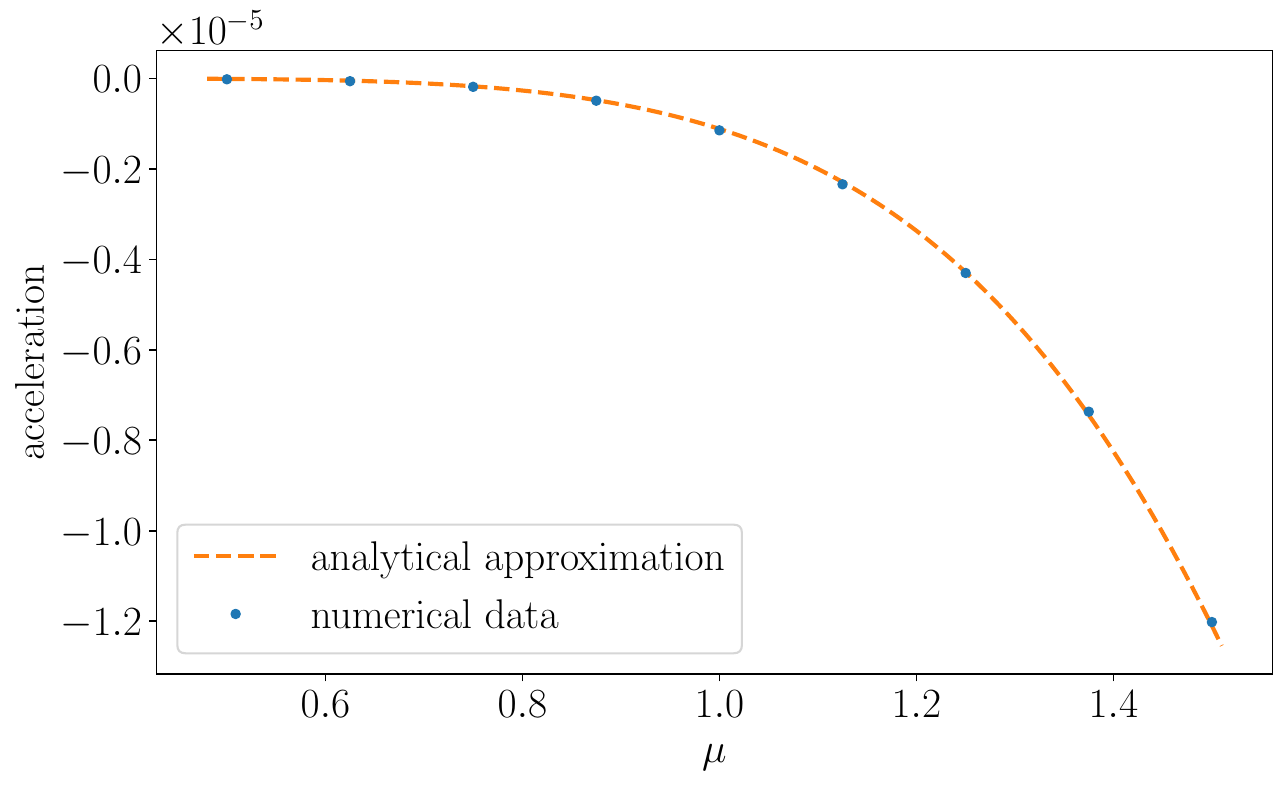}
	\caption{Acceleration of a dark soliton with different values of $\mu$ under the influence of the wave in the second component coming from the left with a frequency $\omega_2^+ = -\tilde{\omega}= 1.4$, $g_{12} = 0.8$ and amplitude $a = 0.05$.}
	\label{fig:scalar_mu_2}
\end{figure}

To verify the above results, we performed numerical simulations. The initial condition was a dark soliton in the first sector and a wave propagating from the left end of the interval with a given frequency and an appropriate wavenumber, where its amplitude was kept sufficiently small. However, the initial wave was multiplied by a superposition of hypebolic tangents to `cut' it smoothly, in order to have the initial wave beginning slightly after the left boundary and ending slightly before the centre of the soliton.  This deviation from a plane wave shape introduces a short `kick' exerted on the soliton, and this results in a constant velocity, on top of which we observe the acceleration compared with Eq.\ (\ref{eq:scalar_second_component_acceleration}). More precisely, the wave in the second component had the form
\begin{equation}
\psi_2(x,t=0) = a e^{i k_2^+ x} \, \Phi_{\text{cut}}(x)\,,
\end{equation}
with parameters such that a wave with $k_2^+$ propagates to the right, i.e., $\tilde{\omega} < 0$ and $\tilde{\omega} < - \mu g_{12}$. $\Phi_{\text{cut}}$ is the 'cutting' function mentioned before:
\begin{equation}
\label{eq:cut}
\Phi_{\text{cut}}(x) = \frac12 \left( \tanh(x - x_{\text{min}} - 10) - \tanh(x + 10) \right)\,,
\end{equation}
where $x_{\text{min}}$ is the left boundary in space. The centre of the soliton was computed as the minimum of the probability density in the first component. Then, the acceleration was computed by fitting the quadratic function to the position of the centre and compared with Eq.\ (\ref{eq:scalar_second_component_acceleration}): see figures \ref{fig:scalar_g_2}, \ref{fig:scalar_frequency_2}, \ref{fig:scalar_amplitude_2} and \ref{fig:scalar_mu_2}. It turned out that our effective linearized model explains the observed accelerations quite well for a wide range of parameters.

\begin{figure}
	\includegraphics[width=\columnwidth]{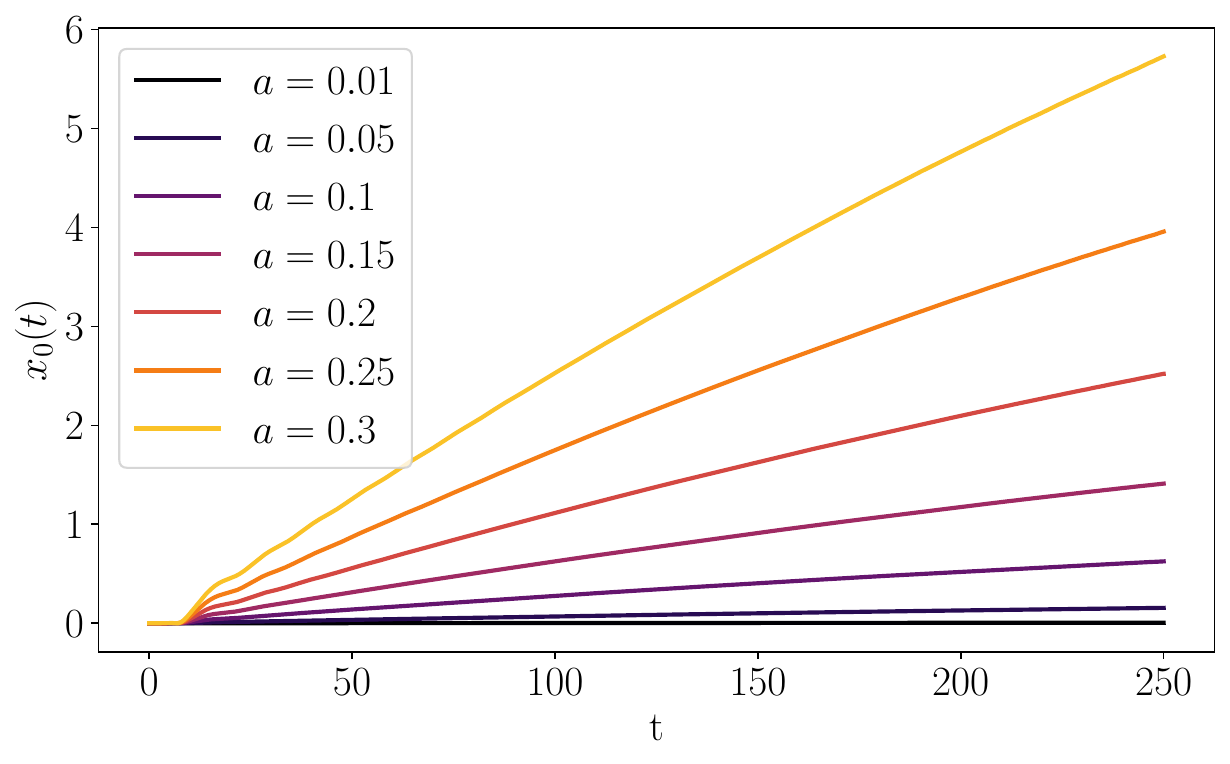}
	\caption{Position of a dark soliton with $\mu = 1$ under the influence of the wave in the second component coming from the left with a frequency $\omega_2^+ = -\tilde{\omega}= 1.4$, $g_{12} = 0.8$ and and different amplitudes.}
	\label{fig:scalar_dynamics_2_amp}
\end{figure}

\subsection{Dark soliton and a wave in the same component}\label{ssec:DS_wave_in_1st_component}
Since the linearised eqs.\ (\ref{eq:linearized}) for $\xi_1$ and $\xi_2$, in the case of the dark soliton, are independent, we start with a similar ansatz as before, namely $\xi_2 = 0$. Now we examine the case where the wave with small amplitude $a$ comes from $-\infty$ in the first component, which (taking into account the solution (\ref{eq:scalar_asymptotic_linearized_1})) has the following asymptotics:
\begin{equation}
\label{eq:scalar_and_first}
\begin{split}
\psi_1(x,t) &\xrightarrow[x \to -\infty]{} a \beta \left( -\frac{\frac{k_1^2}{2}-\tilde{\omega} }{\mu }-1 \right) e^{- i \omega_1^+ t} \left(e^{i k_1 x} + r_1 e^{-i k_1 x} \right) \\
&+ a \beta e^{- i \omega_1^- t} \left(e^{-i k_1 x} + r_1^* e^{i k_1 x}\right) - \sqrt{\mu } e^{-i \mu  t}\,,   \\
\psi_1(x,t) &\xrightarrow[x \to +\infty]{} a \beta \left( -\frac{\frac{k_1^2}{2}-\tilde{\omega} }{\mu }-1 \right) e^{- i \omega_1^+ t}  t_1 e^{i k_1 x}
+ a \beta e^{- i \omega_i^- t} t_1^* e^{-i k_1 x} + \sqrt{\mu } e^{-i \mu  t}\,, \\
\psi_2(x,t) &\xrightarrow[x \to \pm\infty]{} 0\,,
\end{split}
\end{equation}
where
\begin{equation}
\label{eq:first_component_normalization}
\beta = \frac{\sqrt{2} \mu }{\sqrt{\left(k_1^2+2 \mu \right) \left(k_1 \left(\sqrt{k_1^2+4 \mu }+k_1\right)+2 \mu \right)}}\,
\end{equation}
is a normalization constant.
The reflection and transmition coefficients are defined as $R_1 = \left|r_1 \right|^2$ and $T_1 = \left|t_1 \right|^2$ respectively. The coefficient $\beta$ was chosen in such a way as to have the simplest form of the force. From the above asymptotics, the effective force acting on a soliton can be derived analogously as in the previous case, obtaining
\begin{equation}
\label{eq:scalar_first_component_force}
F  = a^2 k_1^2 \left(1 + R_1 - T_1 \right)\,.
\end{equation}

The solutions to the linearized equations (\ref{eq:linearized}) in this case with $\mu = 1$ can be found in \cite{KOVRIZHIN2001392, kuznetsov1988instability, HUANG2008321}. Assuming the asymptotics (\ref{eq:scalar_and_first}) and transforming them for arbitrary $\mu$, we obtain
\begin{equation}
\xi_1^\pm(x) =  \frac{2 \beta k_1}{\left(k_1 \mp 2 i \sqrt{\mu}\right) \left(k_1^2 + 2 \tilde{\omega} \right)} \left[\left(\frac{k_1^2}{2} \mp \tilde{\omega} \right) (1 \pm \frac{2 i \sqrt{\mu}}{k_1} \tanh (x \sqrt{\mu})) + \frac{\mu}{\cosh^2(x \sqrt{\mu})} \right] e^{\pm i k_1 x},
\end{equation}
where $\beta$ is given in (\ref{eq:first_component_normalization}). The solutions imply $R_1=0$, $T_1=1$, and thus no force.
However, we also solved the equations numerically, in order to show the more general method used in the further part of this article. The infinities were approximated by sufficiently large $L$, then $x \in [-L, L]$ (in general the grid is different from the one used in the full PDE simulations of CNLSE). We changed the basis from $(\xi_1^-, {\xi_1^+}^*)$ to the solutions for which the asymptotic form of the equations (\ref{eq:linearized}) is diagonal. Let us denote the solutions in the new basis as $(\tilde{\xi}_1^-, (\tilde{\xi}_1^+)^*)$. Then, the boundary conditions were imposed in the following way:
\begin{equation}
\left. \left( \tilde{\xi}_1^{-'} - i k_1 \tilde{\xi}_1^- \right) \right|_{x = -L} = - 2 i k_1 e^{i k_1 L}\,,
\end{equation}
to have $e^{-i k_1 x}$ at $x = - L$ and
\begin{equation}
\left. \left( \tilde{\xi}_1^{-'} + i k_1 \tilde{\xi}_1^- \right) \right|_{x = L} = 0\,,
\end{equation}
to ensure that there is no $e^{i k_1 x}$ at $x = L$. This corresponds to the incoming wave with $e^{i k_1 x}$ for $\omega_1^+$ moving from left to right. $(\tilde{\xi}_1^+)^*$ describes the non-propagating waves and we imposed analogous boundary conditions on it:
\begin{equation}
\left. \left( \left(\tilde{\xi}_1^+)^*\right)^{'}- i k_{\mathrm{im}} (\tilde{\xi}_1^+)^* \right) \right|_{x = -L} =
\left. \left( \left(\tilde{\xi}_1^+)^*\right)^{'} + i k_{\mathrm{im}} (\tilde{\xi}_1^+)^* \right) \right|_{x = L} = 0\,,
\end{equation}
with  imaginary $k_{\mathrm{im}} = \pm \sqrt{2} i \sqrt{\sqrt{\mu^2 + \tilde{\omega}^2} + \mu}$ instead of $k_1$.  Then, $R_1$ and $T_1$ were computed from the numerical solutions, using the equations
\begin{equation}
\begin{split}
R_1 &= \left| \tilde{\xi}_1^-(-L) - e^{-i k_1 L} \right|^2\,, \\
T_1 &= \left| \tilde{\xi}_1^-(L) \right|^2\,.
\end{split}
\end{equation}
The acceleration, computed from the force (\ref{eq:scalar_first_component_force}) with numerically obtained $R_1$, $T_1$ and the mass derived in the previous section, turned out to be close to zero, as expected.

The above result can be compared with the full PDE simulations of CNLSE. Before we do that, let us discuss a particular difficulty present here. In the case of a scalar dark soliton with a wave in the second component, determining the center of the soliton from numerical data is relatively easy, since the wave and the soliton are in completely different components. Then, the center is simply given by a minimum of the probability density in the soliton component. In general, this is not the case because in many scenarios the waves scatter into both components of the condensate. Therefore, developing the strategy of extracting the position of a soliton from numerical data will be important not only for the waves initially in the first component, but for most other setups as well.

\begin{figure}
	\includegraphics[width=\columnwidth]{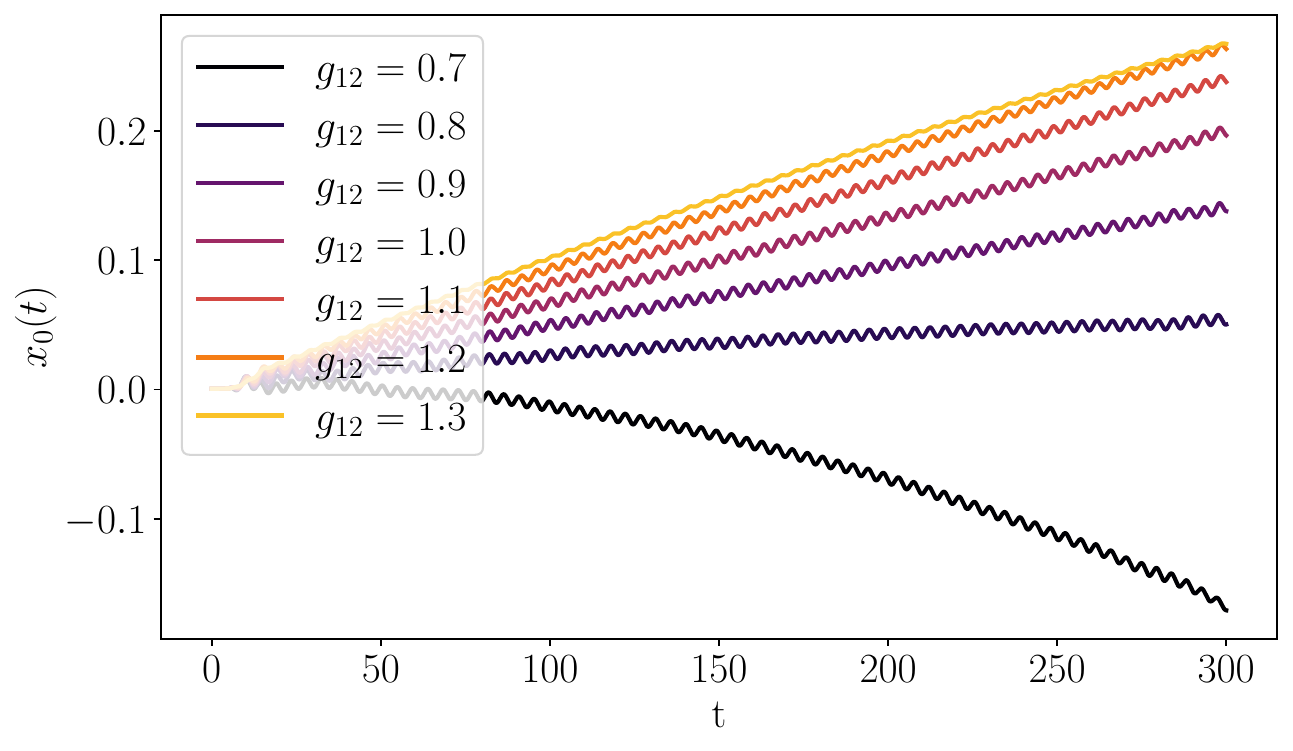}
	\caption{Position of a dark-bright soliton with $\mu = 1$, $\kappa = 0.9$ under the influence of the wave in the bright component coming from the left with frequency $\omega_2^+ = 1.995$ (i.e. $\tilde{\omega} = -1.4$), $a = 0.05$ and different values of $g_{12}$.}
	\label{fig:dark_bright_dynamics_2_g}
\end{figure}

When a wave is present in the same component as the soliton, one can observe oscillations of the minimum (in the case of a bright soliton: maximum) of the probability density.
These oscillations are due to the effect of the incident wave on the soliton, cf.\ Figs.\ \ref{fig:scalar_dynamics_2_amp} and \ref{fig:dark_bright_dynamics_2_g}.
This interesting effect is the analogue described by Quist effect \cite{PhysRevB.60.4240} for vortices. In the present case it
complicates the extraction of the position of the soliton from numerical data.
To improve the determination of the soliton positions, we used filtering of high frequencies from $|\psi(x, t)|^2$ at each instant, $t$, before computing the minimum. Since the amplitude of these oscillations becomes smaller compared to the observed trajectories during the time evolution, it pays off to make time evolution as long as feasible.

\begin{figure}
	\includegraphics[width=\columnwidth]{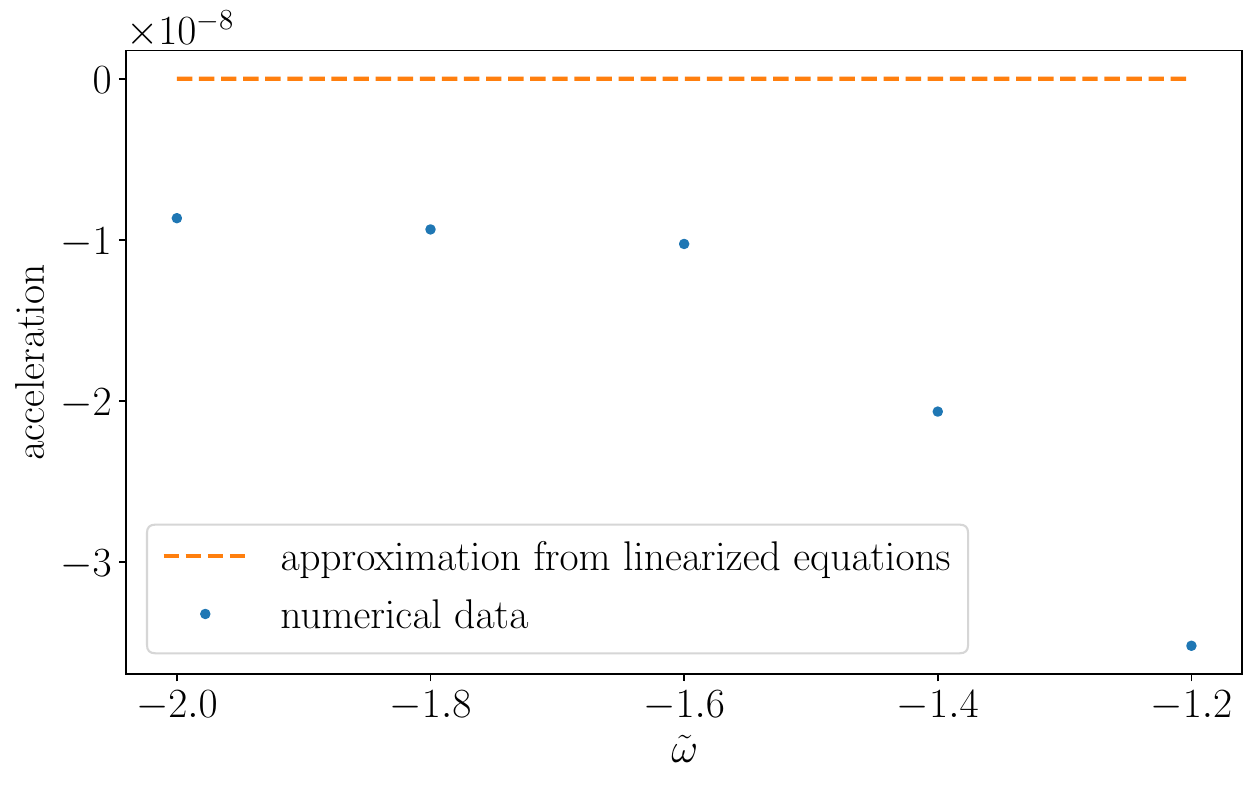}
	\caption{Acceleration of a dark soliton with $\mu = 1$ under the influence of the wave in the first component coming from the left with different frequencies $\omega_1^\pm = \mu \mp \tilde{\omega}$ and amplitude $a = 0.05$.}
	\label{fig:scalar_frequency_1}
\end{figure}

\begin{figure}
	\includegraphics[width=\columnwidth]{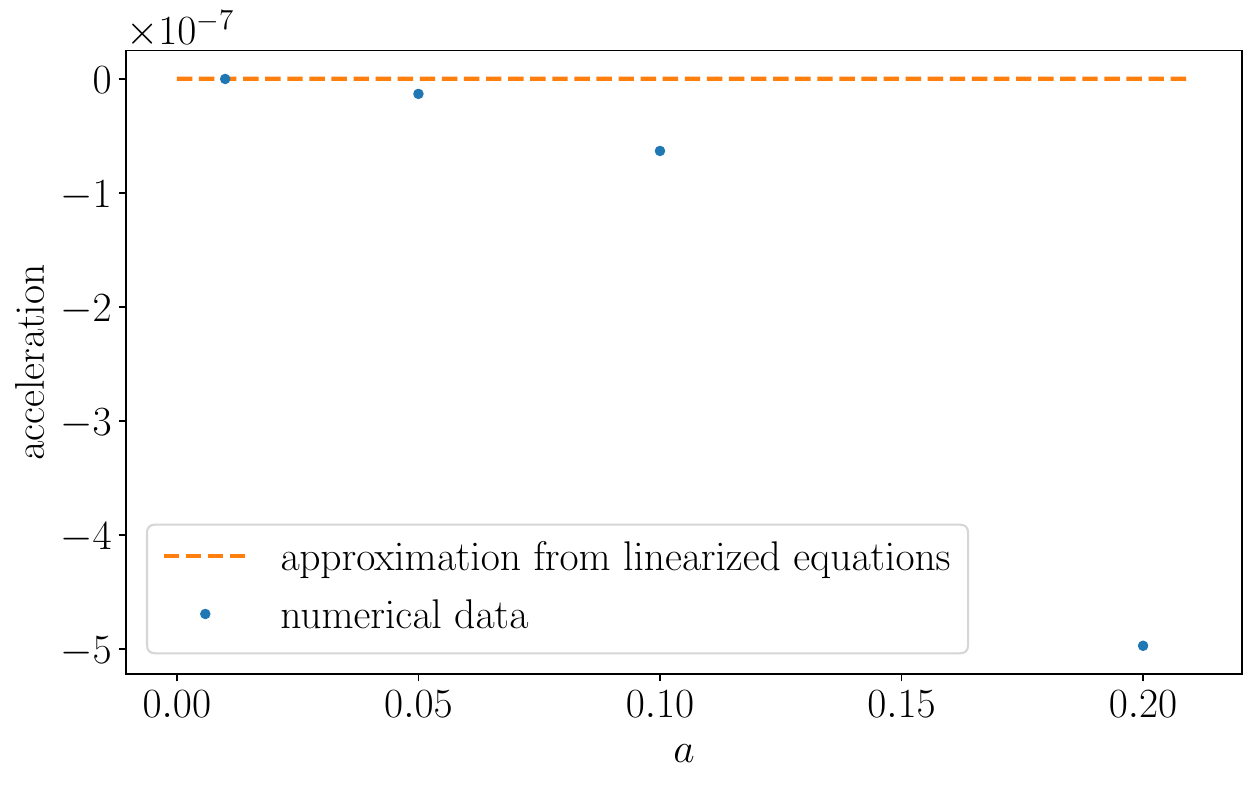}
	\caption{Acceleration of a dark soliton with $\mu = 1$ under the influence of the wave in the first component coming from the left with frequencies $\omega_1^+ = 2.4$, $\omega_1^- = - 0.4$ (i.e. $\tilde{\omega} = -1.4$), $g_{12} = 0.8$ and different amplitudes.}
	\label{fig:scalar_amplitude_1}
\end{figure}

The initial condition in the numerical simulations was $\psi_2(x,t=0) = 0$ and
\begin{equation}
\psi_1(x,t=0) = \Phi(x) + a \beta \left( \left( -\frac{\frac{k_1^2}{2}-\tilde{\omega}}{\mu}-1 \right) e^{i k_1 x}  + e^{-i k_1 x} \right) \, \Phi_{\text{cut}}(x)\,,
\end{equation}
where $\Phi$ is the dark soliton for $t=0$, $\Phi_{\text{cut}}$ is the same `cutting' function as in (\ref{eq:cut}) and $\beta$ is the normalization given by (\ref{eq:first_component_normalization}). We used $\tilde{\omega} < - \mu$ in order to have both $e^{\pm i k_1 x}$ starting a wave moving to the right. Acceleration computed from Eq.\ (\ref{eq:scalar_first_component_force}) divided by the appropriate effective mass with numerically obtained reflection and transmission coefficients has values close to zero. It was compared with the acceleration from the full PDE simulations (figures \ref{fig:scalar_frequency_1} and \ref{fig:scalar_amplitude_1}), which is also small. This seems to indicate that the force is indeed approximately zero.

\section{Dark-bright solitons}\label{sec:dark_bright_soliton}
\subsection{DB solutions}\label{ssec:DB_solutions}
\begin{figure}
\includegraphics[width=\columnwidth]{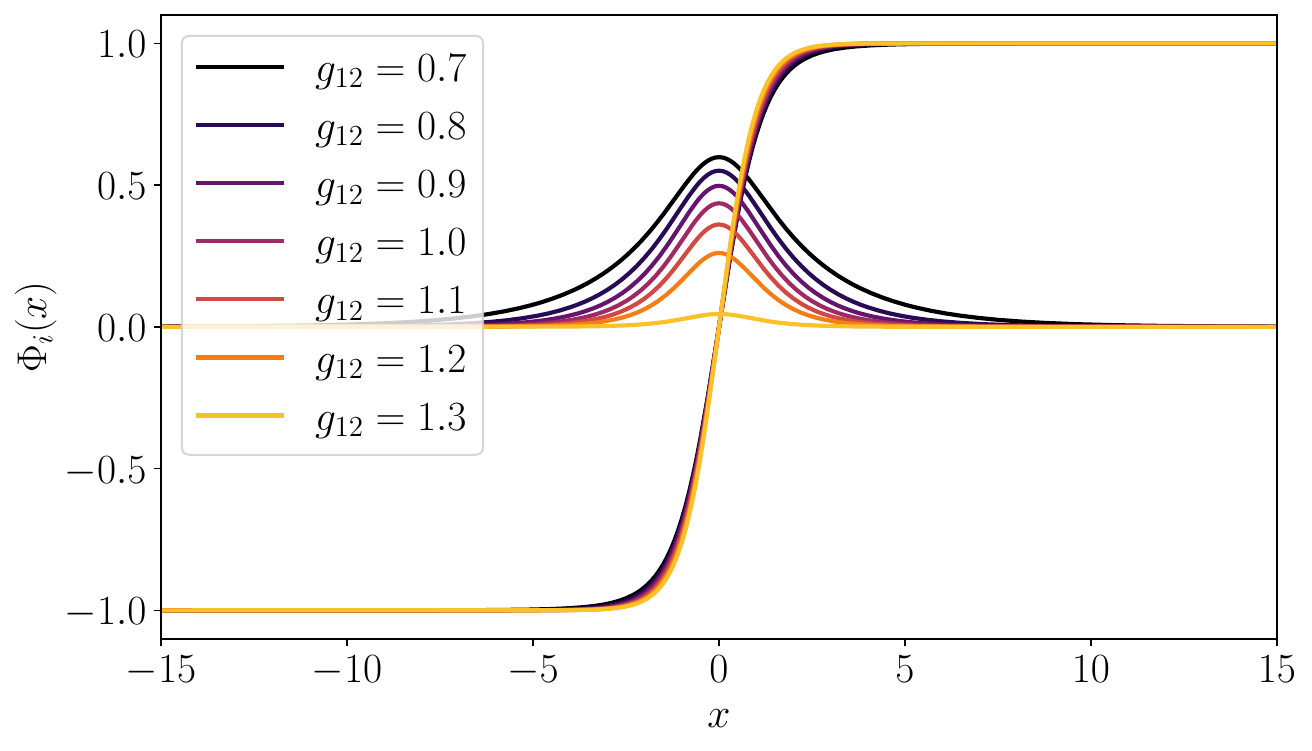}
\caption{Dark-bright solitons for different values of parameter $g_{12}$, calculated using gradient flow method. Parameters used: $\mu = 1$, $\kappa = 0.9$ with $\mu_i$ given as in the Eq.\ (\ref{eq:dark_bright_chemical_potentials}).}
\label{fig:dark_bright}
\end{figure}

The CNLSE equation (\ref{eq:CNLSE}) with $g_{12} = 1$ possesses a particular solution \cite{PhysRevLett.87.010401}
\begin{equation}
\label{eq:dark_bright}
\begin{split}
\psi_1 &=  e^{-i \mu t} \sqrt{\mu} \tanh (\kappa  (x-x_0))\,, \\
\psi_2 &=  e^{-i \left(\mu - \frac{\kappa^2}{2}\right)t} \sqrt{\mu - \kappa^2} \text{sech}(\kappa (x-x_0))\,,
\end{split}
\end{equation}
which is an example of a dark-bright soliton with
\begin{equation}
\label{eq:dark_bright_chemical_potentials}
\mu_1 = \mu\,, \quad
\mu_2 = \mu - \kappa^2/2\,.
\end{equation}
Obviously,  $0 < \kappa^2 < \mu$. Assuming $\psi_i(x,t) = e^{- i \mu_i t} \Phi_i(x)$ we can find DB solitons for other values of the parameter $g_{12}$. Their profiles $\Phi_i$ are presented in fig. \ref{fig:dark_bright}. They can be intuitively understood as a dip in the probability density of atoms of one kind (species or spin state) and a relatively small peak in the probability density of atoms of the other kind.

Consider scattering on such solitons. It can be easily deduced from the Eq.\ (\ref{eq:CNLSE_stationary}) that for $x \to \pm \infty$ DB solitons profiles $\Phi_1(x) \to \pm \sqrt{\mu}$, $\Phi_2(x) \to 0$ and $\Phi_i''(x) \to 0$ regardless of the value of $g_{12}$. Therefore, the matrix $\mathbf{M}$ in Eq.\ (\ref{eq:matrix_M}) becomes asymptotically
\begin{equation}
\mathbf{M}_{x \to \pm \infty} = \mu \left( \begin{matrix}
2 & 1 & 0 & 0 \\
1 & 2 & 0 &  0 \\
0 & 0 & g_{12} & 0 \\
0 & 0 & 0 & g_{12}
\end{matrix} \right)\,.
\end{equation}
Note that this asymptotic form is exactly the same as in the case of scalar dark soliton; therefore, the solutions are the same as in the section \ref{sec:dark_soliton_solution}, in particular the wavenumber
\begin{equation}
k_1 = \sqrt{2} \sqrt{\sqrt{\mu^2 + \tilde{\omega}^2} - \mu}\,,
\end{equation}
and again, the waves in the first component propagate for any $\tilde{\omega} \neq 0$. The only difference is that $\mu_2$ is no longer arbitrary, therefore
\begin{equation}
k_2^\pm = \sqrt{2} \sqrt{\mp \tilde{\omega} - \mu g_{12} + \mu - \kappa^2/2}\,.
\end{equation}
Thus, waves in the second component with wavenumber $k_2^\pm$ propagate when $\mp \tilde{\omega} > \mu g_{12} - \mu + \kappa^2/2$. This means that for $\kappa^2 < 2\mu (1 - g_{12})$ there exists a range of $\tilde{\omega}$ in which both waves can propagate with the same $\tilde{\omega}$, however we were unable to find the stable solitons in this range. For $\kappa^2 \geq 2\mu (1 - g_{12})$ with fixed $\tilde{\omega}$ only one (or neither) of the waves can propagate. Note that this excludes the possibility of propagating both types of waves in the second component for $g_{12} \geq 1$.

Conditions for moving to the right in the first component are the same as for the scalar dark case. In the second component, they are $\tilde{\omega} < \mu - \kappa^2/2$ for $\omega_2^+$ and $\tilde{\omega} > - \mu + \kappa^2/2$ for $\omega_2^-$. Therefore, for $\tilde{\omega} \in (- \mu + \kappa^2/2, \mu - \kappa^2/2)$ both waves propagate to the right (it is always true because $\mu - \kappa^2/2$ is positive).

The moving dark-bright soliton in the Manakov ($g_{12} = 1$) case with velocity $v$ is \cite{PhysRevLett.87.010401}:
\begin{equation}
\label{eq:moving_dark_bright}
\begin{split}
\psi_1 &=  e^{-i \mu t} \sqrt{\mu} \, \left( \cos \alpha \tanh (\tilde{\kappa} (x - x_0 - v t)) + i \sin \alpha \right)\,, \\
\psi_2 &=  e^{-i \left(\mu - \frac{\tilde{\kappa}^2 (1 - \tan^2 \alpha)}{2}\right)t} e^{i v x}
\sqrt{(\mu - \kappa^2) \frac{\tilde{\kappa}}{\kappa}} \, \text{sech}(\tilde{\kappa} (x - x_0 - v t))\,,
\end{split}
\end{equation}
where
\begin{equation}
\tilde{\kappa} = \frac{\kappa^2 - \mu + \sqrt{2\kappa^2 \mu \cos (2\alpha) + \kappa^4 + \mu^2}}{2\kappa}\,,
\end{equation}
\begin{equation}
v = \tilde{\kappa} \tan \alpha\,.
\end{equation}
We compute the renormalized total energy (analogously to the renormalized energy of a dark soliton):
\begin{equation}
E_s = \frac12 \int_{-\infty}^{\infty} \left( |\partial_x \psi_1|^2 + |\partial_x \psi_2|^2 + \left(|\psi_1|^2 - \mu \right)^2 + |\psi_2|^4 + 2 |\psi_1|^2 |\psi_2|^2  \right) {\rm d}x\,,
\end{equation}
and from that we get the effective mass
\begin{equation}
M = \left. \frac{{\rm d}^2 E_s}{{\rm d}v^2} \right|_{v=0}
= \left. \left(\frac{{\rm d}v}{{\rm d}\alpha} \right)^{-2} \left(\frac{{\rm d}^2 E_s}{{\rm d}\alpha^2} - \frac{{\rm d}^2 v}{{\rm d}\alpha^2} \frac{{\rm d}v}{{\rm d}\alpha} \frac{{\rm d} E_s}{{\rm d}\alpha} \right) \right|_{\alpha=0}
= - \frac{2 (\kappa^2 + \mu)}{\kappa}\,.
\end{equation}
For $\kappa \to \sqrt{\mu}$ we reproduce the result for the scalar dark soliton $M = -4\sqrt{\mu}$. Using the renormalized total momentum:
\begin{equation}
P_s = \frac{i}{2} \int_{-\infty}^{\infty} \left(\psi_1 \partial_x \psi_1^*  - \psi_1^* \partial_x \psi_1 + \psi_2 \partial_x \psi_2^*  - \psi_2^* \partial_x \psi_2 \right) {\rm d}x - \mu \Delta \phi\,,
\end{equation}
where $\Delta \phi = 2 \alpha - \pi$ is the phase change between $-\infty$ and $+\infty$ in the dark (first) component, we obtain the same effective mass
\begin{equation}
M = \left. \frac{{\rm d} P_s}{{\rm d}v} \right|_{v=0}
= \left. \left(\frac{{\rm d}v}{{\rm d}\alpha} \right)^{-1} \frac{{\rm d} P_s}{{\rm d}\alpha} \right|_{\alpha=0}
= - \frac{2 (\kappa^2 + \mu)}{\kappa}\,.
\end{equation}
This indicates that the motion of the dark-bright solitons in the Manakov case is indeed Newtonian and that we used the correct renormalization.

\begin{figure}
	\includegraphics[width=\columnwidth]{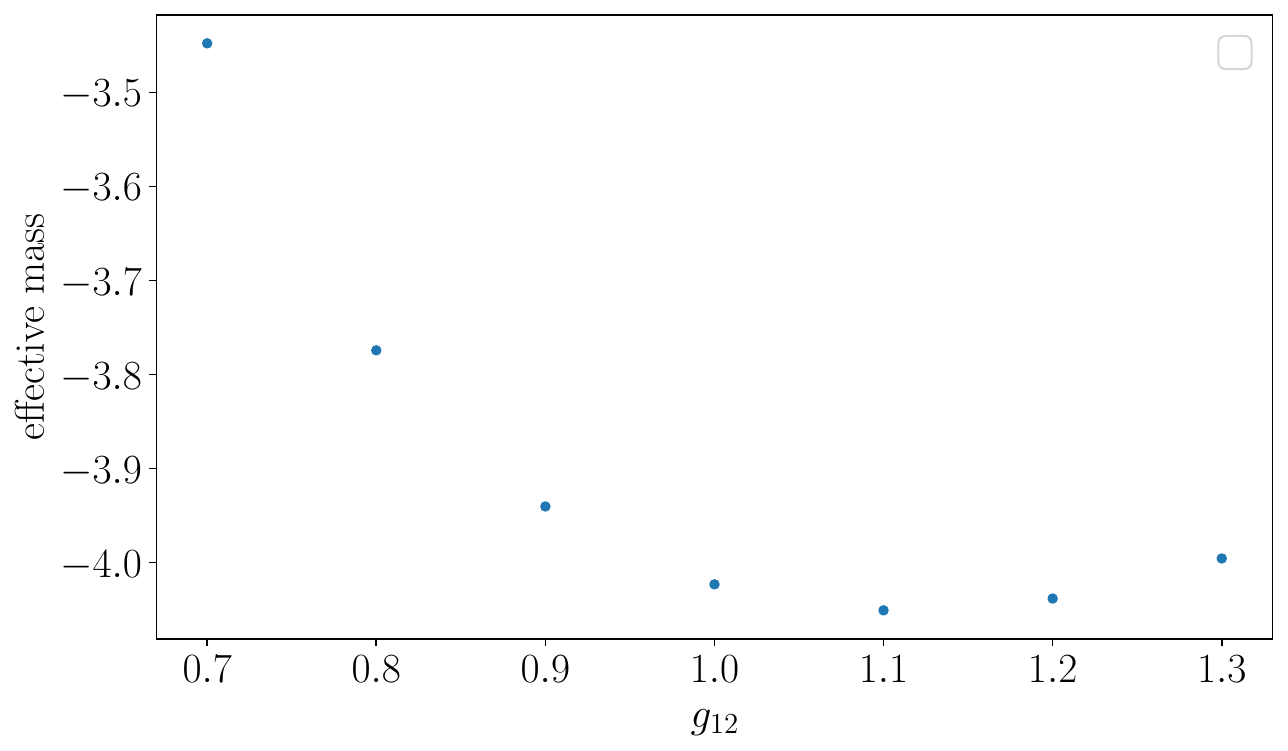}
	\caption{The effective mass of DB solitons for different values of $g_{12}$, obtained using solitons at time $t = 60$ after `pushing' it using the external impulse (\ref{eq:external_impulse}) with parameters: $T = 1$, $V_0 \in [0.005, 0.08]$. Velocities were obtained using the linear fit to the position of the centre of the soliton.}
	\label{fig:dark_bright_mass}
\end{figure}

In the non-Manakov ($g_{12} \neq 1$) case we `push' the soliton using the short and localized external impulse:
\begin{equation}
\label{eq:external_impulse}
V_1(x) = V_0 \, t(T - t) \theta (t) \theta (T-t) \, \frac{\tanh \left(x\right)}{\cosh \left(x\right)}\,,
\quad V_2(x) = 0\,,
\end{equation}
obtain the velocity and moving soliton profile after a sufficiently long time, from which we calculate $E_s$ and compute $M = \left. \frac{d^2 E_s}{dv^2} \right|_{v=0}$ and $M = \left. \frac{d P_s}{dv} \right|_{v=0}$ by fitting a quadratic function to $E_s(v)$ and a linear function to $P_s(v)$, respectively (figure \ref{fig:dark_bright_mass}). Henceforth, we shall use the mass obtained from the momentum, because it is more accurate.

\subsection{Wave in the dark component}
Let us consider a DB soliton with a wave in the first component with a wavenumber $k_1$ and a small amplitude $a$ coming from the left (provided adequate conditions, discussed in the previous section, are met). The asymptotics are then:
\begin{equation}
\label{eq:DB_and_first}
\begin{split}
\psi_1(x,t) &\xrightarrow[x \to -\infty]{} a \beta \left( -\frac{\frac{k_1^2}{2}-\tilde{\omega} }{\mu }-1 \right) e^{- i \omega_1^+ t} \left(e^{i k_1 x} + r_1 e^{-i k_1 x} \right) \\
&+ a \beta e^{- i \omega_1^- t} \left(e^{-i k_1 x} + r_1^* e^{i k_1 x}\right) - \sqrt{\mu } e^{-i \mu  t},   \\
\psi_1(x,t) &\xrightarrow[x \to +\infty]{} a \beta \left( -\frac{\frac{k_1^2}{2}-\tilde{\omega} }{\mu }-1 \right) e^{- i \omega_1^+ t}  t_1 e^{i k_1 x}
+ a \beta e^{- i \omega_i^- t} t_1^* e^{-i k_1 x} + \sqrt{\mu } e^{-i \mu  t}, \\
\psi_2(x,t) &\xrightarrow[x \to -\infty]{} a e^{- i \omega_2^+ t}  r_2^+ e^{-i k_2^+ x}
+ a e^{- i \omega_2^- t}  r_2^- e^{-i k_2^- x},   \\
\psi_2(x,t) &\xrightarrow[x \to +\infty]{} a e^{- i \omega_2^+t}  t_2^+ e^{i k_2^+ x}
+ a e^{- i \omega_2^- t}  t_2^- e^{i k_2^- x},
\end{split}
\end{equation}
where $\beta$ is the same as in Eq.\ (\ref{eq:first_component_normalization}). Using an analogous approach as before, we derive that the force exerted on the soliton by such a wave is
\begin{equation}
\label{eq:DB_first_component_force}
F = a^2 \left(
k_1^2 \left(1 + R_1 - T_1 \right)
+ \left(k_2^+ \right)^2 \left(R_2^+- T_2^+ \right)
+ \left(k_2^- \right)^2 \left(R_2^- - T_2^- \right)
\right),
\end{equation}
where $R_1 = \left|r_1 \right|^2$, $T_1 = \left|t_1 \right|^2$, $R_2^\pm = \left|r_2^\pm \right|^2$ and $T_2^\pm = \left|t_2^\pm \right|^2$.

\begin{figure}
	\includegraphics[width=\columnwidth]{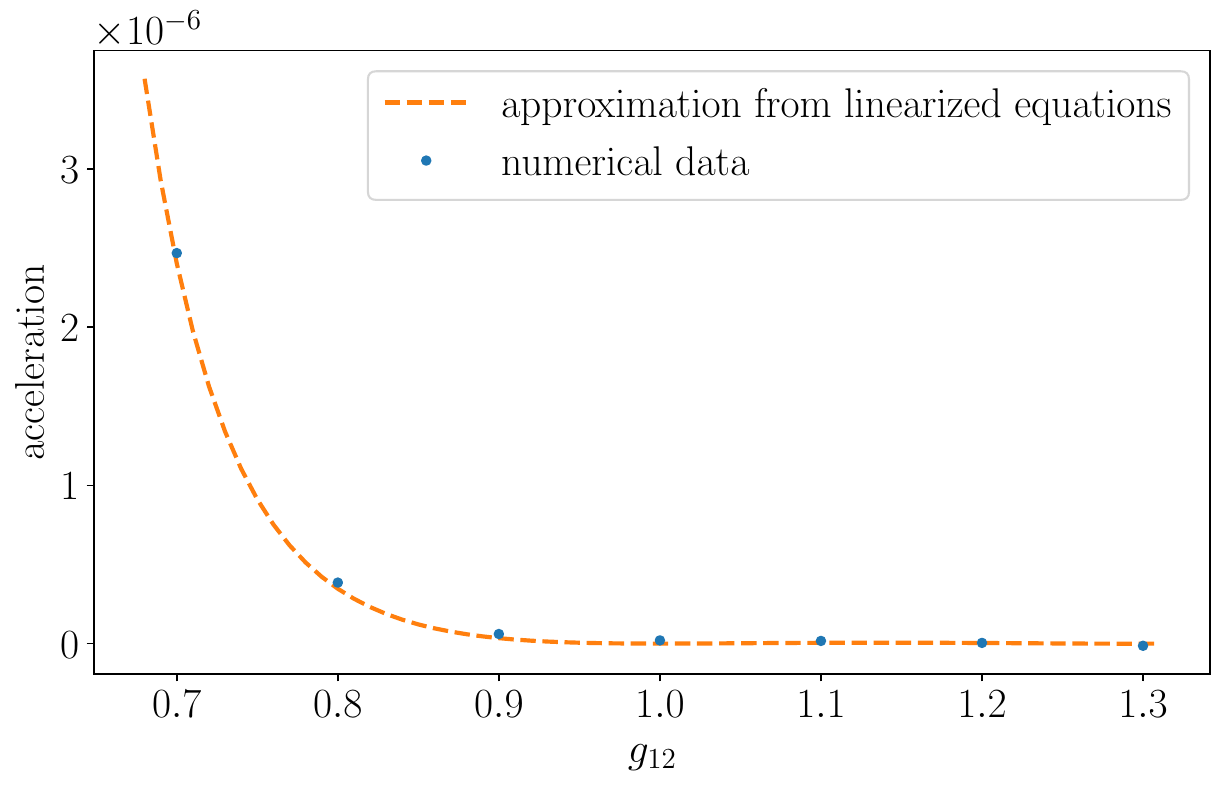}
	\caption{Acceleration of a dark-bright soliton with $\mu = 1$, $\kappa = 0.9$ under the influence of the wave in the dark component coming from the left with frequencies $\omega_1^+ = 2.4$, $\omega_1^- = - 0.4$ (i.e. $\tilde{\omega} = -1.4$), $a = 0.05$ and different values of $g_{12}$.}
	\label{fig:dark_bright_g_1}
\end{figure}

\begin{figure}
	\includegraphics[width=\columnwidth]{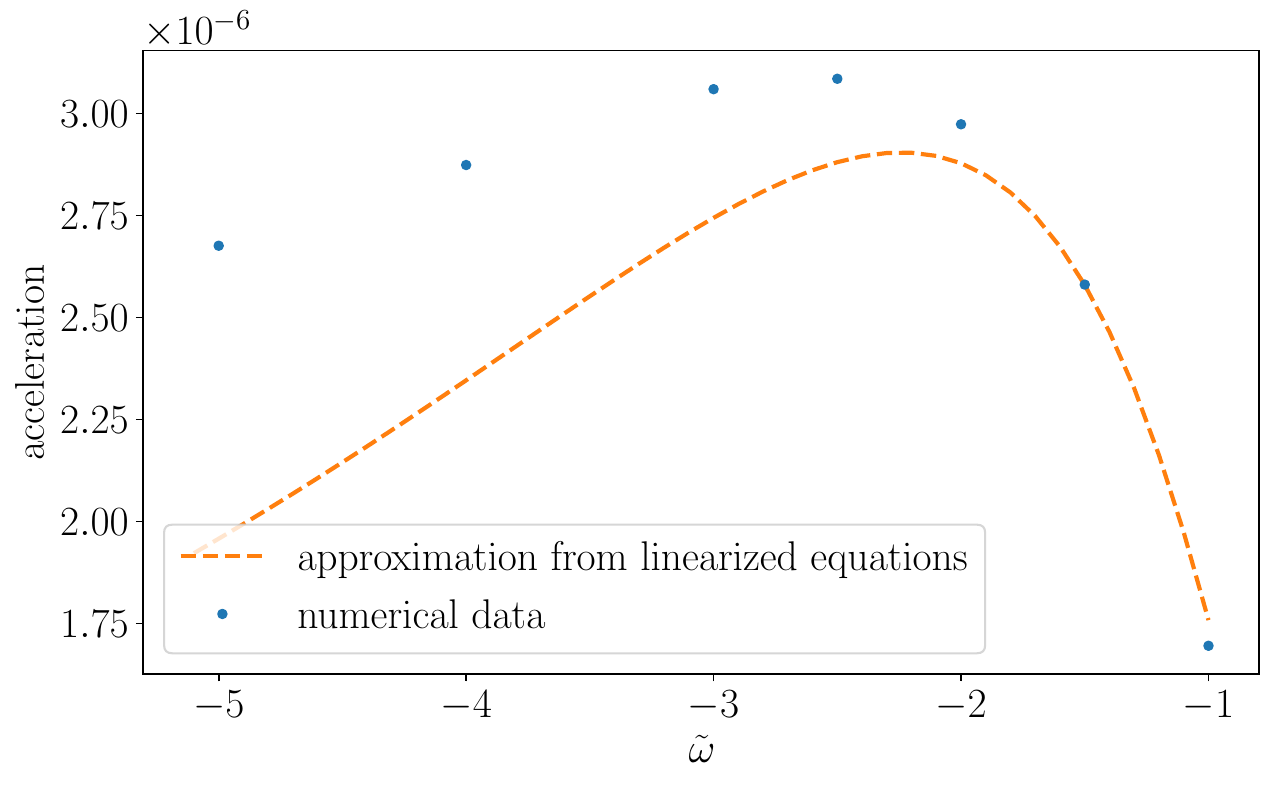}
	\caption{Acceleration of a dark-bright soliton with $\mu = 1$, $\kappa = 0.9$ under the influence of the wave in the dark component coming from the left with an amplitude $a = 0.05$, $g_{12} = 0.7$ and different frequencies $\omega_1^\pm = \mu \mp \tilde{\omega}$.}
	\label{fig:dark_bright_frequency_1}
\end{figure}

\begin{figure}
	\includegraphics[width=\columnwidth]{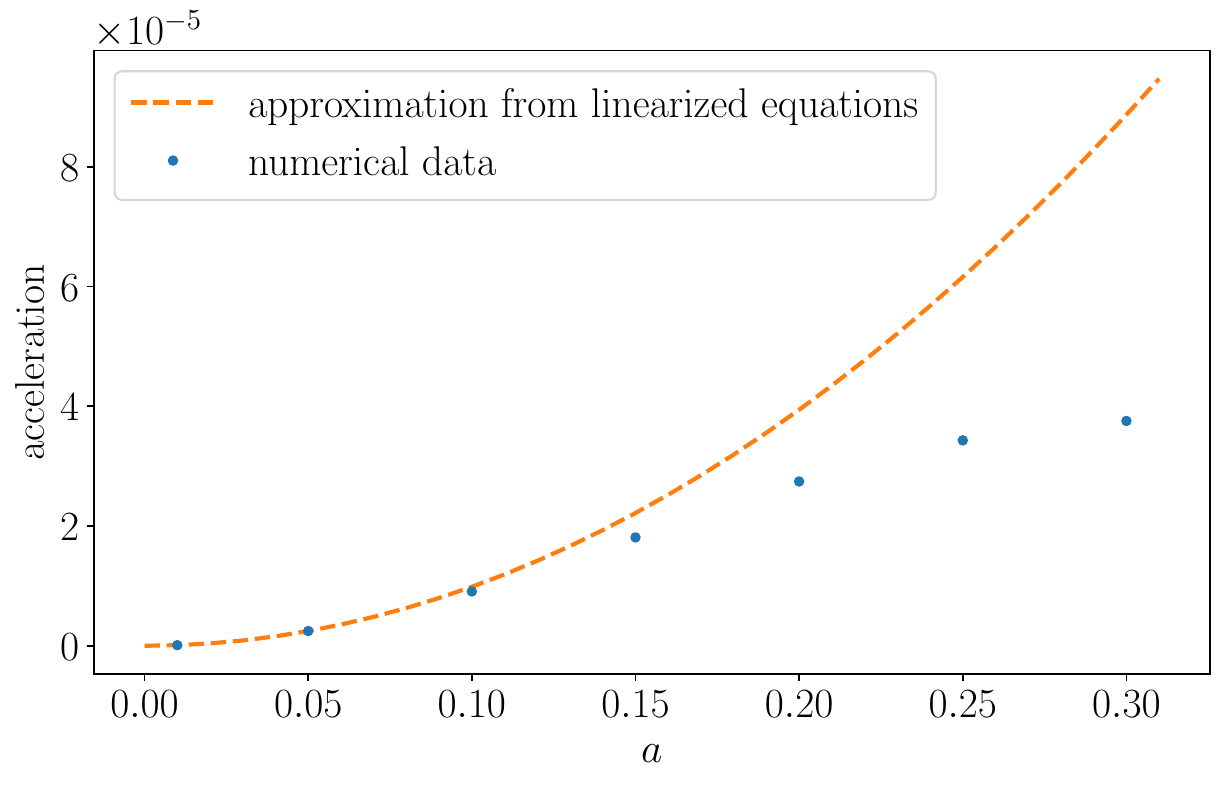}
	\caption{Acceleration of a dark-bright soliton with $\mu = 1$, $\kappa = 0.9$ under the influence of the wave in the dark component coming from the left with frequencies $\omega_1^+ = 2.4$, $\omega_1^- = - 0.4$ (i.e. $\tilde{\omega} = -1.4$), $g_{12} = 0.7$ and different amplitudes.}
	\label{fig:dark_bright_amplitude_1}
\end{figure}

The setup (\ref{eq:DB_and_first}) corresponds to the eigenwave (\ref{eq:scalar_asymptotic_linearized_1}) with $A = 0$, let us call it the first eigenwave. The other eigenwave ($B=0$), i.e. (\ref{eq:DB_and_first}) with $k_1 \to - k_1$ gives the same expression for the effective force, but with different values of reflection and transmission coefficients. Let us focus on the first eigenwave (note that the conditions for propagation to the right are derived above for the first eigenwave). Using a similar approach as for the scalar soliton, we can compute values of these coefficients and (using the effective mass computed above) compare the resulting acceleration with full PDE simulations (with initial field configurations constructed analogously as for the scalar dark soliton). It turns out that we always get the negative radiation pressure, described well by our linear model for relatively small amplitudes and frequencies (figures \ref{fig:dark_bright_g_1}, \ref{fig:dark_bright_frequency_1} and \ref{fig:dark_bright_amplitude_1}). The nonlinear behaviour for larger amplitudes is expected, since linear approach relies on the fact that the amplitude is small. However, the discrepancy for larger frequencies is surprising and requires further study.

\subsection{Wave in the bright component}
If we consider a DB soliton with a wave in the second component with a wavenumber $k_2^\pm$ (and a small amplitude $a$) coming from the left, the asymptotics are
\begin{equation}
\label{eq:DB_and_second}
\begin{split}
\psi_1(x,t) &\xrightarrow[x \to -\infty]{} a \beta \left( -\frac{\frac{k_1^2}{2}-\tilde{\omega} }{\mu }-1 \right) e^{- i \omega_1^+ t}  r_1 e^{-i k_1 x}
+ a \beta e^{- i \omega_1^- t} r_1^* e^{i k_1 x} - \sqrt{\mu } e^{-i \mu  t}\,,   \\
\psi_1(x,t) &\xrightarrow[x \to +\infty]{} a \beta \left( -\frac{\frac{k_1^2}{2}-\tilde{\omega} }{\mu }-1 \right) e^{- i \omega_1^+ t}  t_1 e^{i k_1 x}
+ a \beta e^{- i \omega_i^- t} t_1^* e^{-i k_1 x} + \sqrt{\mu } e^{-i \mu  t}\,, \\
\psi_2(x,t) &\xrightarrow[x \to -\infty]{} a e^{- i \omega_2^\pm t}  e^{i k_2^\pm x}
+ a e^{- i \omega_2^+ t}  r_2^+ e^{-i k_2^+ x}
+ a e^{- i \omega_2^- t}  r_2^- e^{-i k_2^- x} \,,   \\
\psi_2(x,t) &\xrightarrow[x \to +\infty]{} a e^{- i \omega_2^+t}  t_2^+ e^{i k_2^+ x}
+ a e^{- i \omega_2^- t}  t_2^- e^{i k_2^- x}\,,
\end{split}
\end{equation}
with $\beta$ the same as in Eq.\ (\ref{eq:first_component_normalization}). Similarly as before, we can derive the force exerted on the soliton:
\begin{equation}
\label{eq:DB_second_component_force}
F = a^2 \left(
k_1^2 \left(R_1 - T_1 \right)
+ \left(k_2^+ \right)^2 \left(R_2^+- T_2^+ \right)
+ \left(k_2^- \right)^2 \left(R_2^- - T_2^- \right)
+ \left(k_2^\pm \right)^2
\right)\,,
\end{equation}
where $R_1 = \left|r_1 \right|^2$, $T_1 = \left|t_1 \right|^2$, $R_2^\pm = \left|r_2^\pm \right|^2$ and $T_2^\pm = \left|t_2^\pm \right|^2$.

\begin{figure}
	\includegraphics[width=\columnwidth]{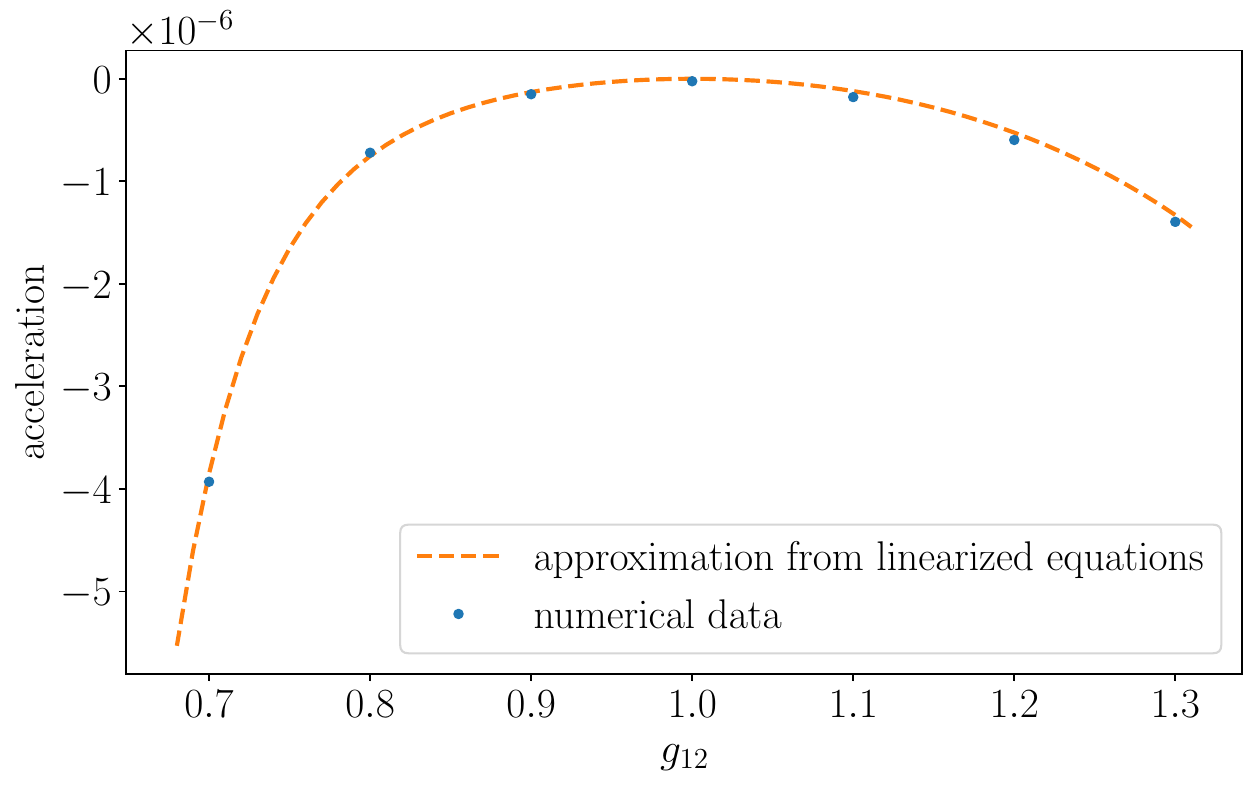}
	\caption{Acceleration of a dark-bright soliton with $\mu = 1$, $\kappa = 0.9$ under the influence of the wave in the bright component coming from the left with frequencies $\omega_2^+ = 1.995$, $\omega_2^- = -0.805$ (i.e. $\tilde{\omega} = -1.4$), $a = 0.05$ and different values of $g_{12}$.}
	\label{fig:dark_bright_g_2}
\end{figure}

\begin{figure}
	\includegraphics[width=\columnwidth]{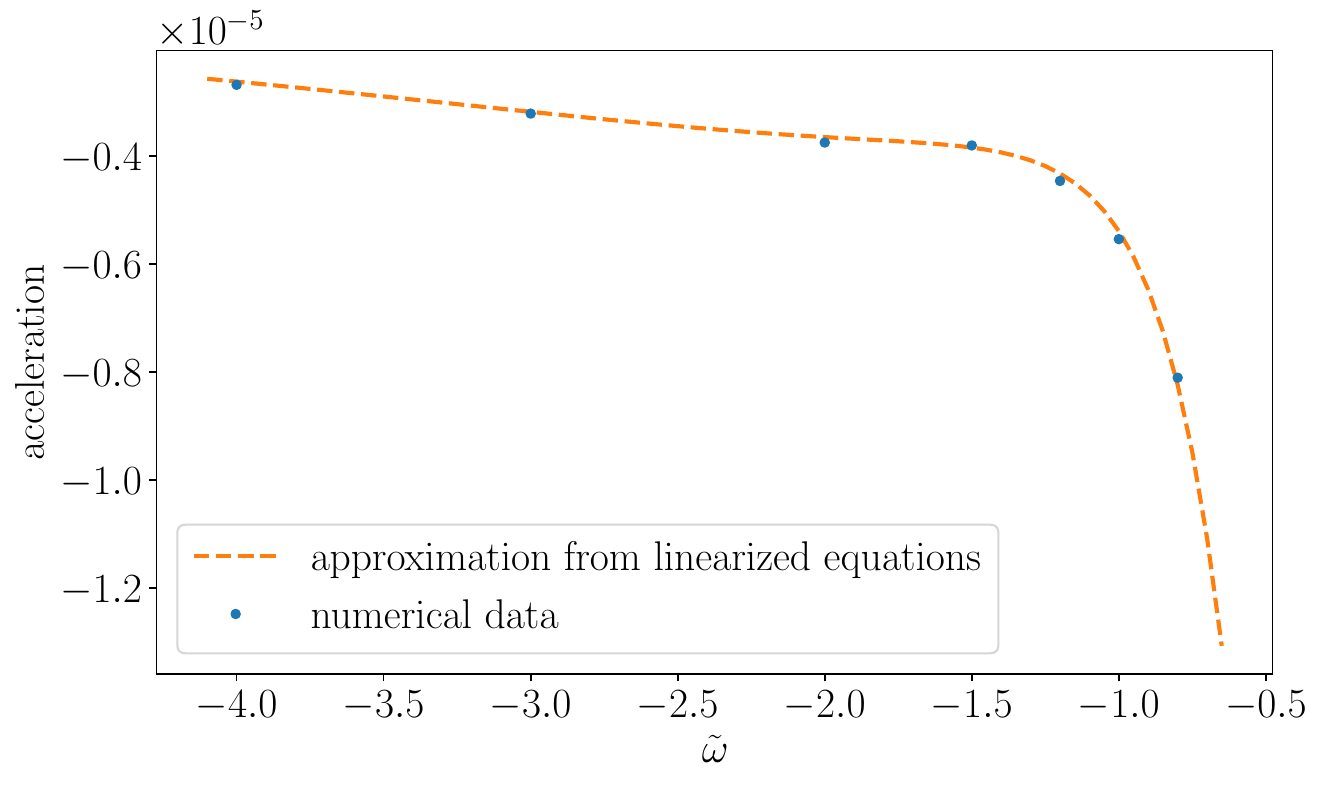}
	\caption{Acceleration of a dark-bright soliton with $\mu = 1$, $\kappa = 0.9$ under the influence of the wave in the bright component coming from the left with an amplitude $a = 0.05$, $g_{12} = 0.7$ and different frequencies $\omega_2^+ = \mu - \kappa^2/2 - \tilde{\omega}$.}
	\label{fig:dark_bright_frequency_2}
\end{figure}

\begin{figure}
	\includegraphics[width=\columnwidth]{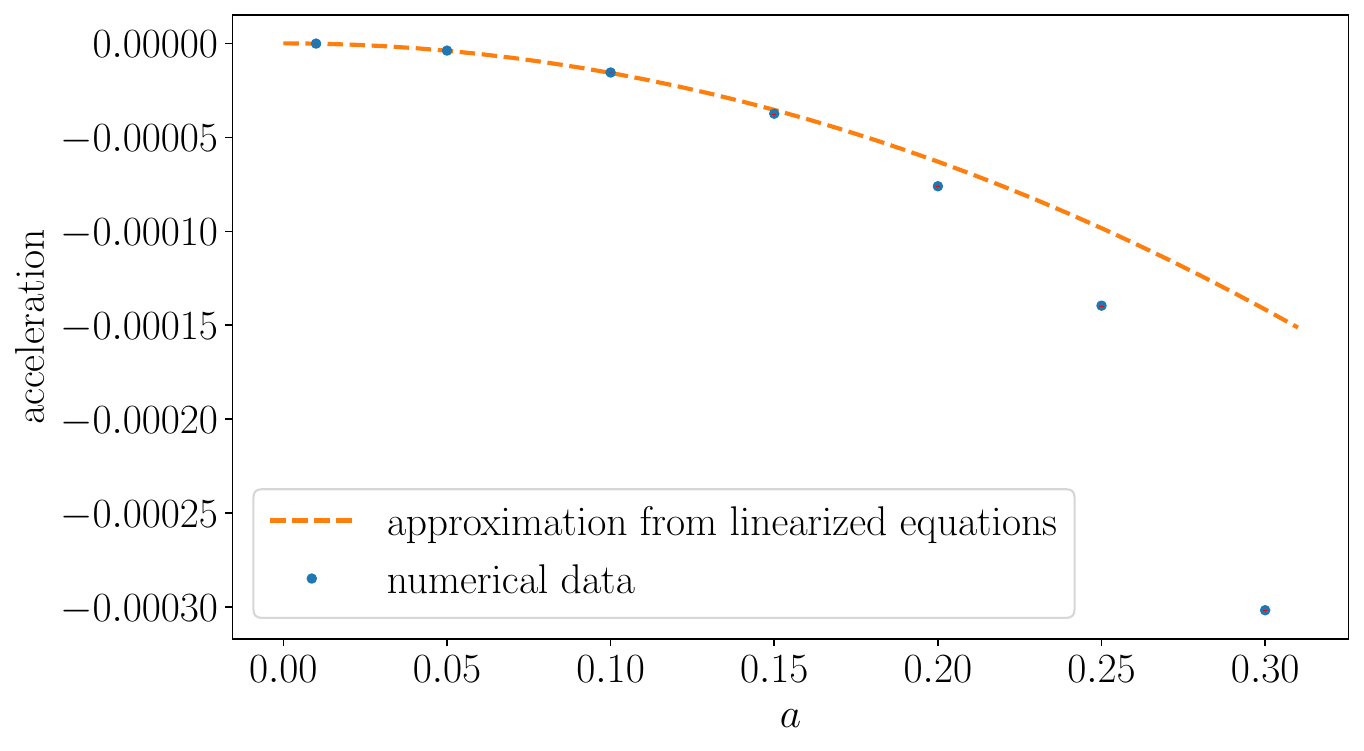}
	\caption{Acceleration of a dark-bright soliton with $\mu = 1$, $\kappa = 0.9$ under the influence of the wave in the bright component coming from the left with frequency $\omega_2^+ = 1.995$ (i.e. $\tilde{\omega} = -1.4$), $g_{12} = 0.7$ and different amplitudes.}
	\label{fig:dark_bright_amplitude_2}
\end{figure}

\begin{figure*}
	\includegraphics[width=\textwidth]{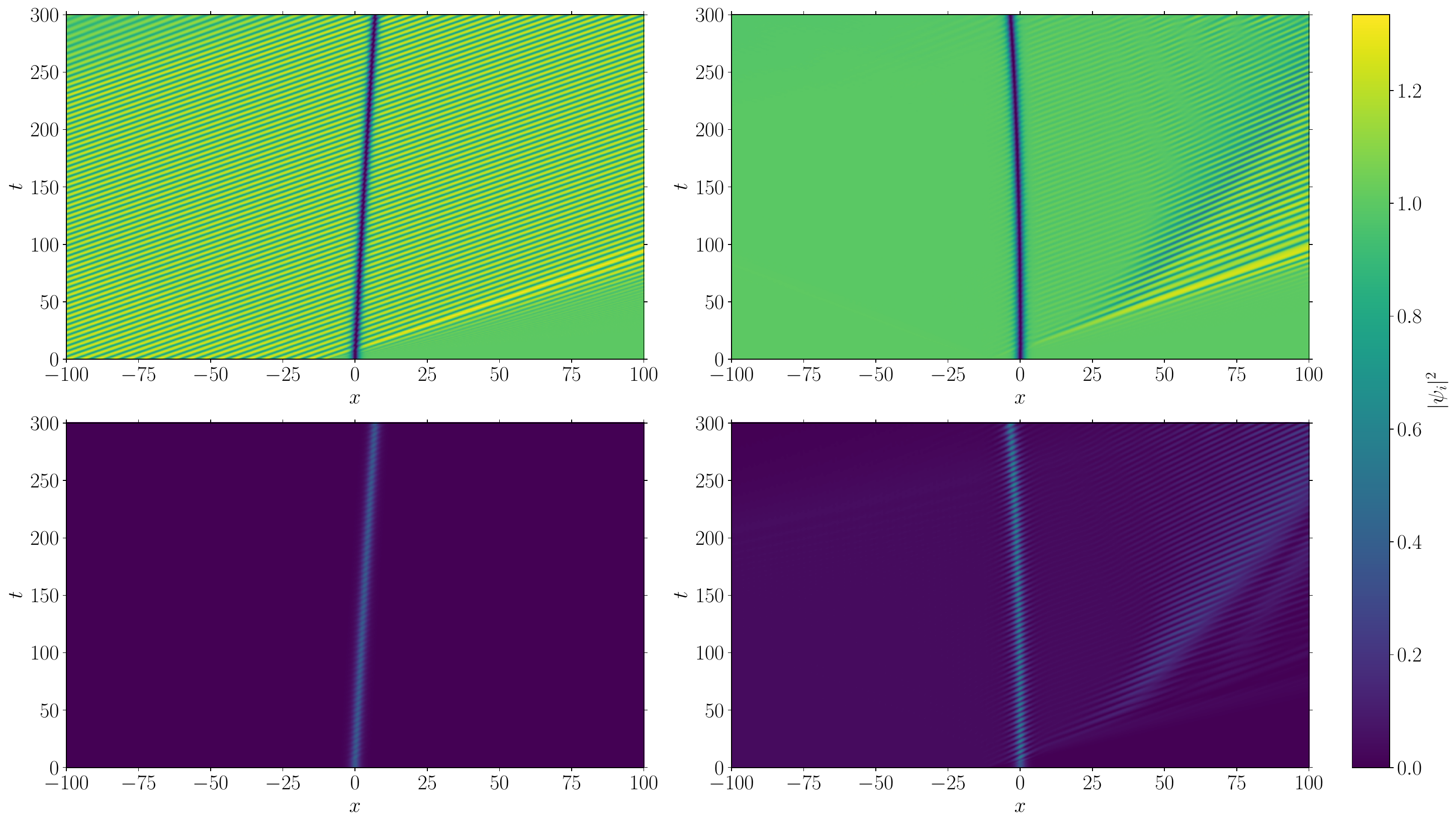}
	\caption{Evolution of the dark-bright soliton with $\mu = 1$, $\kappa = 0.9$, $g_{12} = 0.7$, interacting with a wave in the dark (left) and bright (right) component coming from the left with frequencies corresponding to $\tilde{\omega} = -1.4$ and amplitude $a = 0.2$. The probability densities of the dark (top row) and the bright (bottom row) components are presented.}
	\label{fig:DB_example}
\end{figure*}

Again, we compute the values of the reflection and transmission coefficients numerically and compare the resulting acceleration with the full PDE simulations. In this case we observe the positive radiation pressure for all the values of parameters considered, and everything is described well by the linearized model provided the amplitude is small (figures \ref{fig:dark_bright_g_2}, \ref{fig:dark_bright_frequency_2} and \ref{fig:dark_bright_amplitude_2}). Examples of the evolution for waves in the both components are presented in the figure \ref{fig:DB_example}.

\subsection{DB solitons in a harmonic trap}
\label{sec:trap}
\begin{figure}
	\includegraphics[width=\columnwidth]{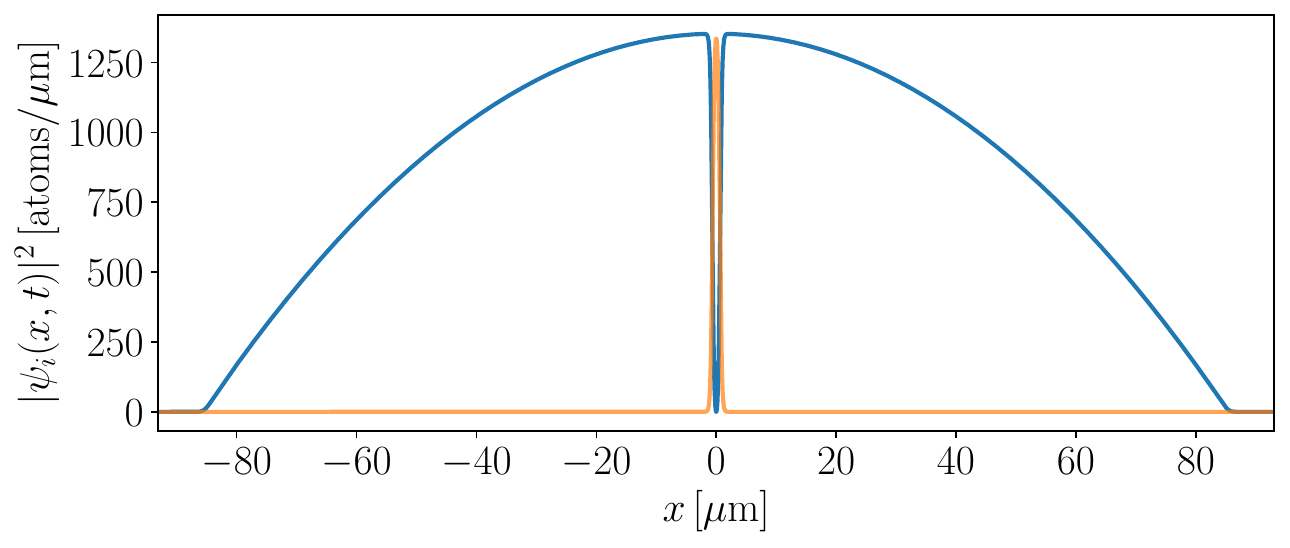}
	\caption{Dark-bright soliton in a harmonic trap with a frequency $\omega_x = 2 \pi \times 14 \mathrm{Hz}$ and $\omega_\perp = 2 \pi \times 425 \mathrm{Hz}$ in a CNLSE with $a_{12} \approx 110 a_0$ (in our units $g_{12} = 1.1$). The number of atoms is about $N_1 \approx 152500$, $N_2 \approx 1700$ in the dark (blue, darker line) and bright (orange, lighter line) component respectively, which corresponds to $\mu = 35000$ and $\kappa = 20$ in our units.}
	\label{fig:dark_bright_V}
\end{figure}

To verify how well the above results apply in realistic situations, let us include a harmonic trapping potential
\begin{equation}
V_1(x) = V_2(x) = \frac12 \omega_x^2 x^2
\end{equation}
in the equation (\ref{eq:CNLSE}), solve it numerically, and compare with the linearized approximations obtained without a trap. In this section, we use physical units, unless we specifically refer to `our units', defined in Sec. \ref{ssec:GPe}. In order for the parameters to be possible to achieve in experiments, we choose the trapping frequencies used in \cite{PhysRevLett.83.5198}, i.e. $\omega_x = 2 \pi \times 14 \mathrm{Hz}$, and $\omega_\perp = 2 \pi \times 425 \mathrm{Hz}$. Similarly, we normalize $\psi_i$ to the number of atoms $N_1 \approx 152500$, $N_2 \approx 1700$ in the dark and bright component respectively. $N_1$ was again inspired by \cite{PhysRevLett.83.5198}, while the proportion $N_2 \approx 0.01 N_1$ is similar to \cite{MIDDELKAMP2011642}. We choose $a_{11} = a_{22} \approx 100 a_0$ and $a_{12} \approx 110 a_0$, where $a_0$ is the Bohr radius. With these parameters, we use the gradient flow method and find a dark-bright soliton, presented in fig. \ref{fig:dark_bright_V}.

\begin{figure}
	\includegraphics[width=\columnwidth]{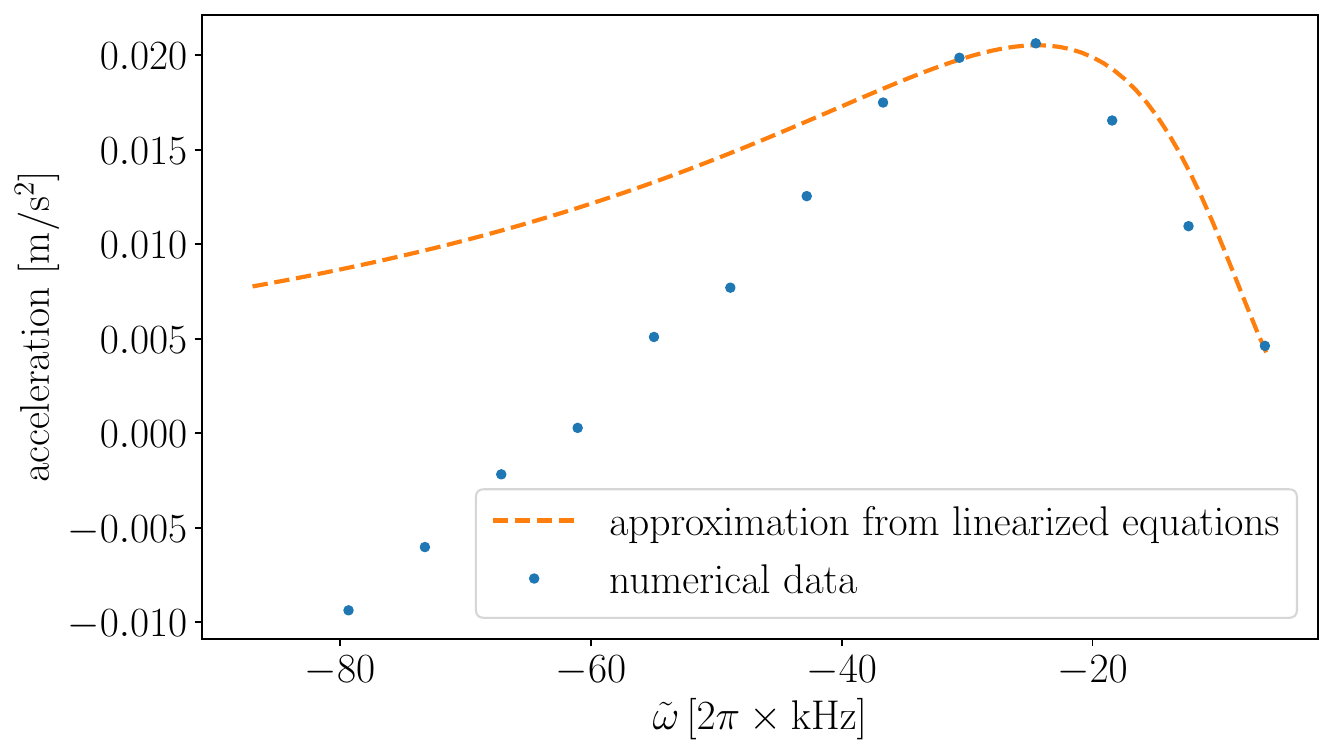}
	\includegraphics[width=\columnwidth]{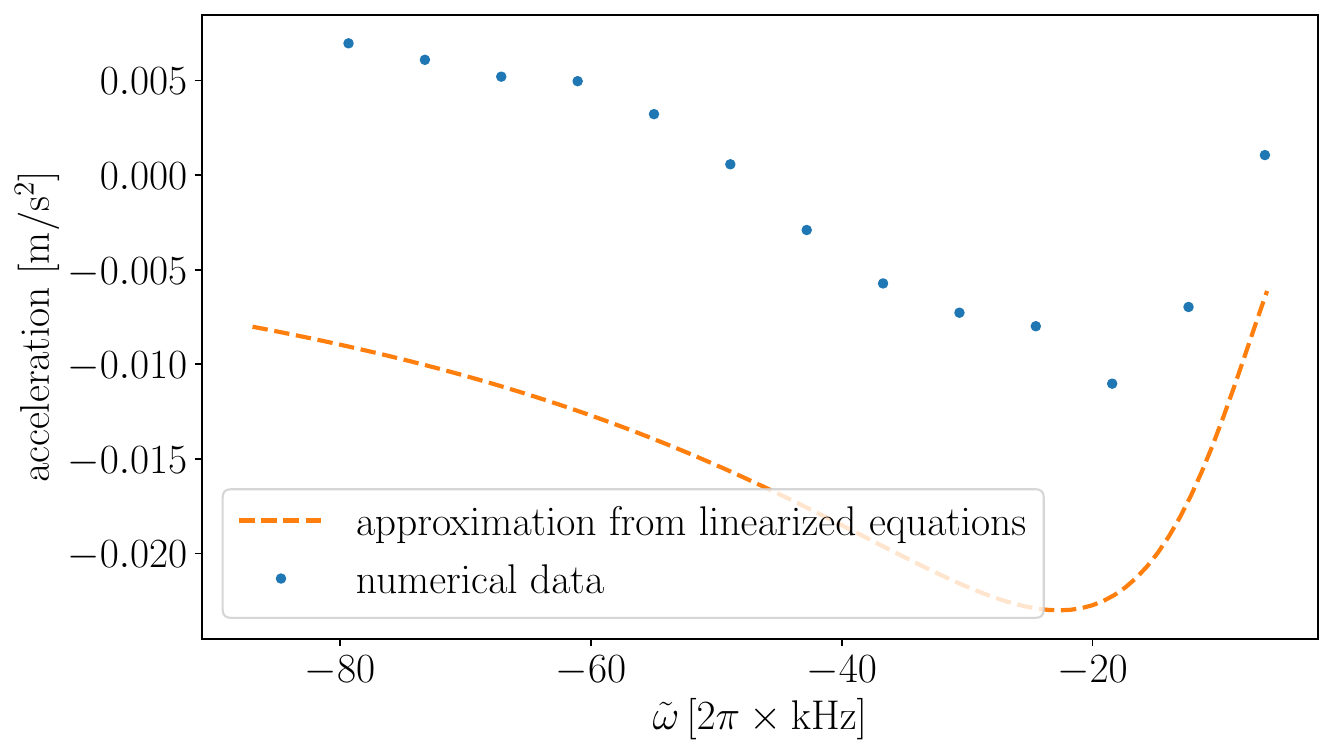}
	\caption{Acceleration of the dark-bright soliton (fig. \ref{fig:dark_bright_V}) in a harmonic trap under the influence of waves in the dark (upper) and bright (lower) component, coming from the left with different frequencies and amplitude $a \approx 0.1 \sqrt{\mu}$, compared with a linear approximation without a trapping potential.}
	\label{fig:dark_bright_frequency_V}
\end{figure}

\begin{figure*}
	\includegraphics[width=\textwidth]{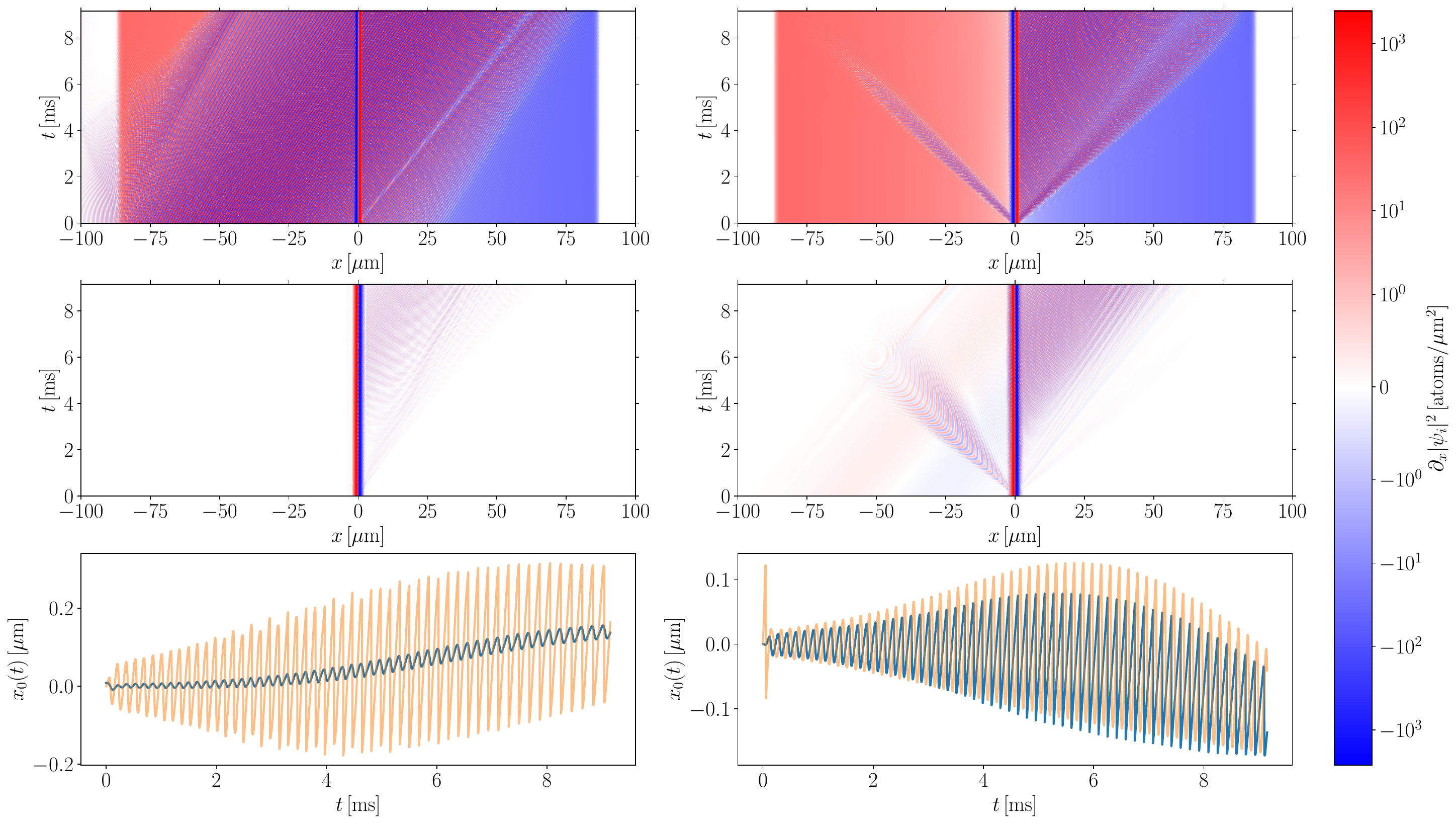}
	\caption{Evolution of the dark-bright soliton in a harmonic trap (fig. \ref{fig:dark_bright_V}), interacting with a wave in the dark (left) and bright (right) component coming from the left with frequencies $\omega_1^+ = 2 \pi \times 12351 \, \mathrm{Hz}$, $\omega_1^- = -2 \pi \times 174 \, \mathrm{Hz}$, $\omega_2^+ = 2 \pi \times 12386 \, \mathrm{Hz}$ (i.e. $\tilde{\omega} = -2 \pi \times 6263 \, \mathrm{Hz}$) and amplitude $a \approx 0.1 \sqrt{\mu}$. Spatial derivatives of the probability density of the dark (top row) and the bright (middle row) components are presented as functions of time and space. Additionally, the positions of the dark (blue, darker line) and bright (orange, lighter line) parts of the soliton are shown as functions of time (bottom row).}
	\label{fig:DB_V_example}
\end{figure*}

We simulated collisions of the DB soliton with monochromatic waves of several frequencies, incoming both from the dark and the bright component. Amplitudes of these waves were equal to about 10\% of the amplitude of the DB soliton's dark component, corresponding to $a = 20$ in our units. Despite the fact, that the linearized approximation does not include a trap, the measured accelerations followed somewhat similar curves (fig. \ref{fig:dark_bright_frequency_V}). The accelerations were measured only in the short time interval (about first 4 ms), since for longer times the force gets affected more by the harmonic potential (see an example of trajectories in fig. \ref{fig:DB_V_example}).

\section{Conclusions}
Using the Newtonian approximation and staying in the linear regime, we have successfully described the acceleration of the scatterer due to the action of the radiation pressure of a wave scattering on a dark and dark-bright soliton. This simple model agrees with numerical simulations for a wide range of parameters. We have shown that a collision of a scalar dark soliton with a wave in the second component of the condensate always results in a positive radiation pressure. For dark-bright solitons, however, we found that the radiation pressure is negative if the wave is incoming from the dark component, and positive otherwise. Our results provide a quantitative description for idealised homogenous BEC, and a qualitative model for more realistic trapped condensates.

The mechanism responsible for NRP in this model relies on the fact that the soliton is present in both components. Otherwise, the equations separate, and from the conservation of energy it follows that the reflection and transmission coefficients sum to one. This implies that the force is always nonnegative, as we have seen explicitly for the scalar soliton case. If the soliton is a vector soliton, then the constraints from the energy conservation allow for both positive and negative sign of the force. Furthermore, the dispersion relations (i.e. the wavenumbers $k_1$ and $k_2^\pm$) play an important role in determining this sign (see table \ref{tab:coefficients}).

\begin{table}
\begin{tabular}{|c|c|c|}
\hline
coefficient & wave from dark & wave from bright \\
\hline
$R_1$ & $1.511 \times 10^{-6}$ & $4.792 \times 10^{-6}$ \\
\hline
$T_1$ & $9.930 \times 10^{-1}$  & $9.325 \times 10^{-3}$ \\
\hline
$R_2^+$ & $2.666 \times 10^{-6}$  & $1.601 \times 10^{-4}$ \\
\hline
$T_2^+$ & $5.188 \times 10^{-3}$  & $9.929 \times 10^{-1}$ \\
\hline
$\delta_{i1} + R_1 - T_1$ & $6.962 \times 10^{-3}$  & $-9.321 \times 10^{-3}$ \\
\hline
$\delta_{i2} + R_2^+ - T_2^+$ & $-5.185 \times 10^{-3}$  & $7.279 \times 10^{-3}$ \\
\hline
$k_1^2 \left(\delta_{i1} + R_1 - T_1 \right)$ & $1.003 \times 10^{-2}$  & $-1.343 \times 10^{-2}$ \\
\hline
$\left(k_2^+ \right)^2 \left(\delta_{i2} + R_2^+ - T_2^+ \right)$ & $-1.343 \times 10^{-2}$  & $1.885 \times 10^{-2}$ \\
\hline
\end{tabular}
\caption{Example of reflection and transmission coefficients (derived from linearized equations) and the formulas present in the effective forces (\ref{eq:DB_first_component_force}) and (\ref{eq:DB_second_component_force}). Kronecker deltas are such that we put $i=1$ for the wave incoming from the dark component, and $i=2$ for the bright. Parameters of the soliton and the scattered wave: $\mu = 1$, $\kappa = 0.9$, $g_{12} = 0.7$, $\tilde{\omega} = -1.4$. Note that the wavenumbers change the sign of the total force.}
\label{tab:coefficients}
\end{table}

The discrepancies between our effective linear model and the full PDE simulations are completely expected for the larger amplitudes of the incoming wave, since the linearization relies on it being small. However, the disagreement for large frequencies of the wave incoming from the dark component and hitting the dark-bright soliton is currently not well understood within the scope of this paper. This could be an opportunity for further research, especially combined with a detailed study of the nonlinear effects, which can play a role here. Another interesting possibility would be to investigate NRP on other solitons in a two-component BEC, such as bright-bright and dark-dark solitons.

It is also worth mentioning, that there is a correspondence between dark/dark-bright solitons and N\'eel/Bloch walls in the parametrically driven nonlinear Schr\"odinger equation \cite{PhysRevD.11.2950, SARKER1976255, PhysRevA.40.3226, PhysRevLett.65.1352, PhysRevE.64.056618, PhysRevE.71.026613, PhysRevE.75.026604, PhysRevE.75.026605}, although with different dynamics. Moreover, the problem described in this paper shows some similarities to the wall-on-wall scattering described in \cite{PhysRevE.75.026605}, which is an interesting topic for futher studies.

To the best of the authors' knowledge, the described setups can be, in principle, reproduced experimentally. Hopefully, in the future, this article could help to promote NRP from being a purely theoretical concept to an observable physical phenomenon.

\begin{acknowledgments}
TR acklowledges the support of National Science Centre, grant number 2019/35/B/ST2/00059. DC and TR thank for the support of the Priority Research Area under the program Excellence Initiative – Research
University at the Jagiellonian University in Krak\'ow. The authors would like to express their gratitude to Adam Wojciechowski and Krzysztof Sacha for useful discussions, especially on the experimental applications of this research.
\end{acknowledgments}

\appendix
\section{Derivation of the total energy and momentum}
\label{sec:integrals_of_motion_derivation}
Using the Lagrangian density (\ref{eq:CNLSE_lagrangian}), one can derive the energy-momentum tensor
\begin{equation}
\tensor{T}{^\mu_\nu} = \sum_{i=1,2} \left( \frac{\partial \mathcal{L}}{\partial (\partial_\mu \psi_i)} \partial_\nu \psi_i + \frac{\partial \mathcal{L}}{\partial (\partial_\mu \psi_i^*)} \partial_\nu \psi_i^* \right) - \mathcal{L} \delta_\nu^\mu\,,
\end{equation}
where $\mu, \nu = 0, 1$, $\partial_0 = \partial_t$ and $\partial_1 = \partial_x$. Since we consider CNLSE without the trapping potential, the Lagrangian density (\ref{eq:CNLSE_lagrangian}) is invariant under translations, and then from the Noether theorem follows that such a tensor is a conserved current, meaning that it obeys
\begin{equation}
\label{eq:energy_momentum_conservation}
\sum_{\mu=1,2} \partial_\mu \tensor{T}{^\mu_\nu} = 0\,.
\end{equation}
The total energy and momentum are defined, respectively:
\begin{equation}
E = \int_{-\infty}^{\infty}\tensor{T}{^0_0} {\rm d}x, \quad
P = -\int_{-\infty}^{\infty}\tensor{T}{^0_1} {\rm d}x.
\end{equation}
Computing the energy-momentum tensor and integrating, we obtain their explicit form (\ref{eq:total_energy_explicit}) and (\ref{eq:total_momentum_explicit}). Then, equations (\ref{eq:energy_momentum_conservation_explicit_E_total}) and (\ref{eq:energy_momentum_conservation_explicit_P_total}) follow from (\ref{eq:energy_momentum_conservation}).

\section{Numerical methods}
\label{sec:numerical_methods}
In all of the simulations of soliton dynamics in the full PDE we have used the second order split-step method \cite{10.1007/978-3-642-18775-9_64}. This method requires periodic boundary conditions; therefore, at large $x$ the dark soliton (and the dark part of the dark bright soliton) was `glued' with the antisoliton to achieve $\psi_1 = -\sqrt{\mu}$ at the right boundary. The spatial step was $\Delta x = 0.1$, while the temporal step was in the range from $\Delta t = 0.0001$ to $\Delta t = 0.0003$, depending on a particular simulated configuration. We used $x \in [-500, 1000]$ or $x \in [-1000, 2000]$. Other methods with other boudary conditions are possible to implement in CNLSE, see e.g. \cite{doi:10.1137/130920046}.

Linearized equations were solved using sparse matrices. The derivatives were discretized using the five-point stencil. The grid was $x \in [-20, 20]$ with the step $\Delta x = 0.01$. The size of the grid in this problem does not need to match the size of the spatial grid used in solving the full PDE. In fact, we verified that the solutions to linearized equations do not depend on the grid size, provided that it is sufficiently large, such that the solutions achieve the expected asymptotic form.

\bibliography{CNLSE_article_bibliography}

\end{document}